\documentclass[preprint,aps,draft,linenumbers,showpacs,aps]{revtex4}

\usepackage{amsfonts}
\usepackage{amsmath}
\usepackage{bm}
\usepackage{graphicx}  
\usepackage{color}
\usepackage{mathrsfs}

\input epsf 

\usepackage{changes}
\definechangesauthor[name={Oreste}, color=red]{OP}
\definechangesauthor[name={Franco}, color=blue]{FV}
\definechangesauthor[name={Ciccio}, color=green]{FM}

\begin{document}


\title{Exact hybrid-Vlasov equilibria for sheared plasmas with in-plane and out-of-plane magnetic field}

\author{F. Malara, O. Pezzi and F. Valentini}

\affiliation{Dipartimento di Fisica, Universit\`a della Calabria, 87036, Rende (CS), Italy} 

\begin{abstract} 
The hybrid Vlasov-Maxwell system of equations is suitable to describe a magnetized plasma at scales of the order of or larger than proton 
kinetic scales. An exact stationary solution is presented by revisiting previous results with a uniform-density shear flow, directed either 
parallel or perpendicular to a uniform magnetic field, and by adapting the solution to the hybrid Vlasov-Maxwell model. A quantitative 
characterization of the equilibrium distribution function is provided by studying both analytically and numerically the temperature 
anisotropy and gyrotropy and the heat flux. In both cases, in the shear region, the velocity distribution significantly departs from local 
thermodynamical equilibrium. A comparison between the time behavior of the usual ``fluid-like'' equilibrium shifted Maxwellian and the exact 
stationary solutions is carried out by means of numerical simulations of the hybrid Vlasov-Maxwell equations. These hybrid equilibria can be 
employed as an unperturbed states for numerous problems which involve sheared flows, such as the wave propagation in inhomogeneous 
background and the onset of the Kelvin-Helmholtz instability.
\end{abstract}

\maketitle 

\section{Introduction}

Shearing flows in plasma are found in many natural contexts, like, for instance, the interface between the solar wind and planetary 
magnetospheres (e.g., \cite{fujimoto98,hasegawa04,sckopke81}), the interaction region between fast and slow streams of the solar 
wind \cite{brunocarbone13}, or astrophysical jets \cite{hamlin13}. 

A magnetized plasma configuration with a shearing flow is stable if the jump $\Delta u$ of the bulk velocity ${\bf u}$ accross 
the shear is lower than a certain threshold, which is typically of the order of the background Alfv\'en velocity component parallel to 
${\bf u}$ \cite{chandrasekhar61}. In this case, waves possibly propagating in the plasma are affected by the velocity shear 
through different effects. In particular, small scales can be progressively generated in the wave pattern in the direction of the bulk 
velocity gradient by a phase-mixing mechanism; this effect is similar to phase-mixing of Alfv\'en waves which propagate in a static 
background with an inhomogeneous Alfv\'en speed profile (e.g., \cite{heyvaerts83, pucci14}). The formation of structures at increasingly smaller scales can locally change the nature of waves, for instance converting an Alfv\'en wave into a Kinetic Alfv\'en wave 
\cite{vasconez15, pucci16, valentini17}. Finally, the inhomonegeity associated with the shear couples propagating modes, generating energy 
transfers among different kind of waves \cite{pucci16}.

In the opposite case of sufficiently large $\Delta u$ (or small parallel magnetic field), a shearing flow in a plasma is unstable 
and undergoes the Kelvin-Helmholtz (KH) instability \cite{chandrasekhar61} which leads to the formation of vortices located at the shear layer. The dynamics 
can be more complex in configurations where one component of the magnetic field changes sign across the shear layer 
\cite{hofmann75, einaudi86, chen90, faganello10}, which generates an interplay between KH instability and magnetic reconnection. 
When the width of the shear layer becomes of the order of the ion inertial 
length, dispersive effects can modify the properties of the KH instability \cite{chacon03}. 
The presence of KH vortices has been revealed all along the flanks of the 
low-latitude Earth magnetopause \cite{sckopke81,chen93,kivelson95,fairfield00} (see also \cite{faganello17} for a review of the KH 
instability in the magnetosphere context). Moreover, the development of a mixed KH-tearing mode instability has been considered to take 
place in cometary plasma tails \cite{malara89}. 
A large Reynolds numbers, KH instability can become a source of turbulence through 
nonlinear interactions among vortices or secondary instabilities \cite{matsumoto04, nakamura04, matsumoto06, matsumoto07, faganello08, 
henry13, rossi15}. As an obervational example, we can mention a turbulent layer observed in the Earth's magnetopause \cite{dimare18}, 
where KH vortices forms in consequence of the solar wind motion relative to the magnetosphere. 

From a theoretical point of view, prior to the study of either wave evolution or KH instability in shearing-flow plasma, 
there is the problem of setting up a stationary configuration with a shear flow, where the above-mentioned phenomena can occur. 
This topic does not present any difficulty in 
the framework of the Magnetohydrodynamics (MHD), where a large variety of stationary flows can be envisaged, provided that a null total 
force acts on the magnetofluid at all positions. In contrast, the situation is totally different in the framework of a kinetic theory. In 
fact, building up a stationary particle distribution function (DF) representing a shear flow in the presence of a background magnetic 
field represents a non-trivial problem. Such an issue is particularly relevant in contexts where the width of the shear layer is of the 
order of typical kinetic scales, such as the ion inertial length or Larmor radius. Indeed, in these cases, kinetic aspects become relevant 
and can affect some properties of the stationary configuration, like profiles of the bulk flow, temperature, heat flux, etc., modifying 
them with respect to what can be deduced from fluid approaches. For instance, this is the typical situation which is encountered 
when one tries to model the shear flow at the Earth's magnetopause \cite{faganello17}.

The problem of deducing a stationary configuration with a shear flow for a magnetized plasma within a kinetic description has received a 
certain attention in the literature. In the case of uniform magnetic field, Cai et al. \cite{cai90} have deduced a set of stationary ion 
distribution functions giving origin to 1D shear flows directed perpendicularly to the magnetic field. In particular, they have pointed out 
a different behaviour that is found according to the orientation of the flow vorticity with respect to the magnetic field. A similar topic 
has been treated by other authors: in the uniform ${\bf B}$ case, a form for a stationary shearing kinetic configuration has been deduced 
within the problem of calculating properties of ion-cyclotron modes \cite{ganguli88,nishikawa88}; the same problem has been considered 
in the non-uniform ${\bf B}$ case, deriving a set of specific profiles for bulk velocity and magnetic field \cite{mahajan00}. 
More recently, a form of stationary distribution function has been considered in the case of a non-uniform parallel magnetic field and 
bulk velocity \cite{roytershteyn08}. In the above approaches the fully kinetic problem has been considered, calculating the distribution 
functions of both ions and electrons. 

Despite of the existence of such results, in many studies of the KH instability a shifted local Maxwellian distribution has been 
considered as unperturbed plasma state (e.g., Ref. \cite{karimabadi13}). A shifted Maxwellian has the advantage to be more 
manageable because it allows to easily choose profiles of density, bulk velocity, temperature and magnetic field. On the other hand, at 
shear scales comparables with ion kinetic scales a shifted Maxwellian does not represent a stationary state: in fact, initializing the 
system with a shifted Maxwellian leads to the generation of oscillations \cite{cerri13}, whose amplitude becomes larger when decreasing the scale 
of the shear. Though this aspect could be considered as not relevant in the formation of the final turbulent state, it could somehow 
affect the linear stage of the instability. Moreover, a shifted Maxwellian does not describe the differences which arise in shears with a 
different vorticity orientation with respect to the magnetic field \cite{cai90}, as happens in the dusk or dawn flanks of the 
magnetopause, where the vorticity to magnetic field orientation is reversed. 

More importantly, when studing wave propagation in shearing flows, the use of exact stationary states instead of a shifted Maxwellian is 
crucial. In fact, spurious oscillations associated with a non-exact stationary state would superpose to waves making difficult 
to single out effects due to waves.

When one is interested in describing phenomena at scales comparable with ion scales, a 
successful approach is represented by the so-called hybrid Vlasov-Maxwell (HVM) model, in which ions are 
kinetically described while electrons are treated as a massless fluid \cite{valentini07}. It represents a sort of compromise between the 
usual ``MHD-like'' coarse-grained fluid description and an exceedingly complex fully kinetic approach. The HVM model has been 
adopted for describing several phenomena occurring at scales where the kinetic ion physics starts to play a significant role into the 
plasma dynamics \cite{servidio12, matthaeus14, vasconez15, franci15, servidio15, valentini16, cerri16, pezzi17a}. 
Within the HVM framework, a method to derive ion distribution functions representing approximated kinetic stationary solutions 
has been recently presented by Cerri et al. \cite{cerri13}. It is based on the ``extended fluid model approach'', which considers finite 
Larmor radius effects in the determination of the ion pressure tensor, and it has been recently adopted to describe temperature 
anisotropies due to the shear velocity \cite{cerri14,delsarto16}. However, it should be noticed that the solutions proposed by Cerri et al. 
\cite{cerri13} are not exactly stationary, even in the framework of the HVM theory. In fact, oscillations are still present, even if with 
amplitudes definitely smaller than those recovered in the shifted Maxwellian case. 

In the present paper we derive exact stationary solutions for the set of the HVM equations describing a 
magnetized plasma with an arbitrary shearing flow ${\bf u}$ profile in two different configurations, namely, with a uniform magnetic 
field ${\bf B}$ parallel or perpendicular to ${\bf u}$. The derivation of our solutions is inspired by full-kinetic 
solutions previously obtained in analogous configurations \cite{ganguli88,nishikawa88,roytershteyn08}, which have been adapted to 
the HVM model. In particular, an explicit analytical expression for the solution is found in the parallel ${\bf B}$ case, while 
in the case of perpendicular ${\bf B}$ the solution is calculated by a numerical procedure which integrates single-particle trajectories.
The interest of these solution is twofold: first, they are exactly stationary, thus can be safely used as unperturbed states either in 
wave propagation models and in instability studies; second, due to the properties of the HVM model, they realistically represents 
situations in which the width of shear layer is of the order of or larger than ion kinetic scales, avoiding the complexity of a 
fully kinetic treatment.

The plan of the paper is the following: in Section II we introduce the equations of the HVM model; in Sections III and IV we derive 
and discuss the stationary solution in the cases of parallel and perpendicular magnetic field, respectively; in Section V we present the results of numerical simulations where we analyze the behaviour of the solution in comparison with that of a shifted Maxwellian; finally, in Section VI we summarize the results. \\


\section{Equations of the model}
We consider a quasi-neutral magnetized plasma composed of kinetic protons and a massless fluid of isothermal electrons (the current 
analysis easily extends to heavier ions). We are interested in describing shears occurring at scales larger or comparable with proton 
kinetic 
scales: i.e. $l\gtrsim d_p \sim \rho_{_{Lp}}$ and $\tau \gtrsim \Omega_{cp}^{-1}$, being $d_p=V_{_A}/\Omega_{cp}$ the proton skin depth;
$\rho_{_{Lp}}=v_{th,p}/\Omega_{cp}$ the proton Larmor radius; $\Omega_{cp}=e B_0/m_p c$ the proton cyclotron frequency; 
$V_{_A}=B_0/\sqrt{4\pi n_0m_p}$ the Alfv\'en speed; $v_{th,p}=(k_B T_0/m_p)^{1/2}$ the proton thermal speed. Note that $\beta_p=2 
v_{th,p}^2/V_{_A}^2$; $m_p$, $e$, $n_0$ and $T_0$ are respectively the proton mass, charge, density and temperature; while $B_0$, $c$ and 
$k_B$ are the magnetic field typical value, the light speed and the Boltzmann constant. 

At these scales, the proton dynamics is successfully modeled by the hybrid Vlasov-Maxwell (HVM) equations: 
\begin{eqnarray}
& & \frac{\partial f}{\partial t} + {\bf v}\cdot \nabla f + \frac{e}{m_p}\left( {\bf E} +\frac{\bf v}{c}\times {\bf B} \right) \cdot
\frac{\partial f}{{\bf \partial v}} =0 \label{vlasoveq} \\
& & \frac{\partial {\bf B}}{\partial t}= -c \nabla \times {\bf E} \; \; \; ; \; \; \; {\bf j}=\frac{c}{4\pi} \nabla \times {\bf B} 
\label{FarAmp} \\
& & {\bf E}=-\frac{1}{c}\left( {\bf u}\times {\bf B} \right)+ \frac{1}{en} \left( \frac{{\bf j}\times {\bf B}}{c} \right) -
\frac{1}{en} \nabla p_e \label{E}
\end{eqnarray}
where $f=f({\bf x},{\bf v},t)$ is the proton distribution function in the phase space $( {\bf x}, {\bf v})$. The electric ${\bf E}({\bf 
x},t)$ and magnetic ${\bf B}({\bf x},t)$ fields are respectively determined by the generalized Ohm's law and by the Faraday and Ampere laws, 
by neglecting the displacement current. In Eqs. (\ref{vlasoveq}--\ref{E}), $n({\bf x},t) = \int d^3{\bf v} f({\bf x},{\bf v},t)$ is the 
proton number density, ${\bf u} ({\bf x},t) = \int d^3{\bf v}\, {\bf v} f({\bf x},{\bf v},t)/n({\bf x},t)$ is the proton bulk speed 
and ${\bf j}({\bf x},t)$ is the current density. Electrons are a massless fluid, whose density is equal to that of ions $n_e=n$ for the 
quasi-neutrality condition. The electron pressure $p_e$ is determined by imposing a closure assumption for the electron dynamics (such as 
isothermal or adiabatic state equation). 

Our goal is to build up a stationary hybrid equilibrium state for a sheared flow in presence of a homogeneous background magnetic field. The 
shear is directed along $y$ and depends on $x$, i.e. ${\bf u_0}=u_0(x){\bf e}_y$, and spatial variations occur only along $x$. Two different 
cases are discussed: i) magnetic field ${\bf B}$ {\it parallel} case to ${\bf u_0}$, i.e. $ {\bf B}=B_0{\bf e}_y $; ii) magnetic field 
${\bf B}$ {\it perpendicular} to ${\bf u_0}$, i.e. $ {\bf B}=B_0{\bf e}_z $. These two cases will be investigated separately in the 
following of the paper. 

In the parallel case, no electric field is needed to set the initial equilibrium and electrons are a massless isothermal fluid: $T_e={\rm 
const} \rightarrow p_e = k_B n T_{e}$. In the perpendicular case, an equilibrium electric field, 
directed along $x$, is needed to equilibrate the ${\bf u_0}\times {\bf B}$ term in Eq. (\ref{E}): ${\bf E}= E(x) {\bf e}_x$. In this case we 
also need to relax the electrons closure, by treating the electron pressure $p_e$ as a further independent quantity determined by the 
following equation:
\begin{equation}\label{eadiab}
\left[ \frac{\partial}{\partial t} + \left({\bf u}_e \cdot \nabla\right) \right] \left( \frac{p_e}{n^{\gamma_e}} \right) = 0
\end{equation}
where $\gamma_e$ is the electron adiabatic index. Last equation implies that the electron temperature is not homogeneous. Therefore:
\begin{equation}
 E(x) = - \frac{u_0 B_0}{c} - \frac{1}{n e}\frac{d p_e}{dx}
\end{equation}
The presence of such electric field introduces a charge separation, which - in principle - is not taken into account within the HVM model 
($n_e=n$). However, the discrepancy from the quasi-neutrality condition is extremely small.

To conclude this section we remark that the first attempt for modeling a plasma configuration with a velocity shear at kinetic scales 
essentially extends the fluid, large scales, equilibrium at smaller scales (e.g., Ref. \cite{karimabadi13}). 
These equilibria, based on the local thermodynamical hypothesis \cite{chandrasekhar61}, assume the following form for the proton VDF 
(sheared Maxwellian):
\begin{equation}\label{PDFatt}
f_{SM}(x,{\bf v},t) =  \frac{n_0}{(2\pi)^{3/2} v_{th,p}^3}\exp \left[ -\frac{ v_x^2 + \left( v_y -u_0(x)\right)^2 + v_z^2}{2 
v_{th,p}^2} \right]
\end{equation}
where ${\bf v}=(v_x, v_y, v_z)$, $n_0$ is the proton density and ${\bf u_0}=u_0(x) {\bf e}_y$ is the bulk velocity, being $u_0(x)$ a given 
function.  It can be easily verified that $f_{SM}$ is not a stationary solution of HVM 
equations in both cases (parallel and perpendicular) discussed above.

In the following of the paper, we will revisit and formalize, for the HVM equations, the derivation of kinetic stationary equilibria
for a sheared flow, by considering two different geometrical configurations. We anticipate that our approach is easier with respect to the 
ones adopted in previous works \cite{cai90,ganguli88,nishikawa88,mahajan00,roytershteyn08} and implementing our equilibrium in the HVM 
code is hence quite simple. This allows 
to perform hybrid kinetic simulations with a ``proper'' (and simple) hybrid kinetic equilibrium. We also remark that, the sheared Maxwellian 
DF is often adopted also for analyzing phenomena occurring at kinetic scales in sheared flows, such as KHI. This choice can be easily 
justified for the investigation of a particular class of phenomena, where nonlinearities play a crucial role in developing turbulence, such 
as KHI (e.g., \cite{karimabadi13}). One can correctly argue that, for these processes, the final difference that would occur starting with the 
sheared Maxwellian or with the ``proper'' equilibrium DF is minimal and does not affect the final dynamical state. However, for another 
class of processes (such as phase mixing of a linear wave in a velocity shear \cite{vasconez15,pucci16,valentini16}), starting from a 
correct equilibrium is crucial for correctly investigating the phenomenon itself. In this direction, several works have been focused on the 
extension the MHD-like fluid equilibrium to the microphysics \cite{cerri13,cerri14,delsarto16,delsarto18} and our work gives a 
further contribution in this direction.

\section{Stationary solution for the parallel case: ${\bf u_0}\, ||\, {\bf B}$}
\label{sect:parEQDF}

Here we revisit the derivation of the stationary solution of Eqs. (\ref{vlasoveq}--\ref{E}), in the case in which velocity shear and 
homogeneous magnetic field are parallel and the electric field is vanishing. The method we adopt is similar to the Harris approach for 
deriving kinetic equilibria corresponding to localized current sheets \cite{harris62} and it is based on the determination of the proton DF 
as a function of the motion constants, derived from the Lagrangian description of single-particle dynamics. A similar derivation can be 
found in Ref. \cite{roytershteyn08} in the framework of fully kinetic theory; here, we reconsider the same problem to adapt the solution 
to the HVM model. Moreover, we discuss in a deeper detail the properties of the derived stationary DF, illustrating the spatial
profiles of its moments (bulk velocity, temperatures, heat flux).

For the present geometry, the Lagrangian of the single-particle dynamics is:
\begin{equation}
 \mathcal{L} = \frac{m_p}{2} \left( v_x^2 + v_y^2 + v_z^2\right) - \frac{e}{m_p} v_zB_0 x
\end{equation}
where ${\bf A}=A_z {\bf e}_z = B_0 x\, {\bf e}_z$ is the vector potential associated with the magnetic field. From the integrals of 
motion of $\mathcal{L}$, which are the generalized momenta $P_y$, $P_z$ and the energy $E$, we can define three auxiliary constants:
\begin{equation}
 \begin{cases}
    &k_1 = P_y =m_p v_y \\
    &k_2 = - \displaystyle{\frac{P_z c}{e B_0}} = x - \displaystyle{\frac{v_z}{\Omega_{cp}}} \\
    &k_3 = E = \frac{m_p}{2} \left( v_x^2 + v_y^2 + v_z^2\right)
  \end{cases}
\label{eq:k123}
\end{equation}
To write the proton DF $f_{eq,||}$, we consider the following combinations of the above constants, having the dimension of a velocity, $ 
\alpha_1 = \left\{ \frac{2}{m_p} \left[ E - \frac{P_y^2}{2 m_p}\right] \right\}^{1/2}$ and $\alpha_2 = \frac{P_y}{m_p} - U\left(x - 
\frac{v_z}{\Omega_{cp}}\right)$, being $U=U(k_2)$ the arbitrary shear, function of the motion integral $k_2$. 

Since $\alpha_1$ and $\alpha_2$ are motion integrals, each generic function $F\left(\alpha_1^2 + \alpha_2^2\right)$ is a stationary 
solution of the Vlasov equation \cite{krall73}. The function $F$ is determined by imposing that $f_{eq,||}$ reduces to a Maxwellian with 
density $n_0$ and thermal speed $v_{th,p}$ in absence of the shear ($U=0$), hence:
\begin{equation}\label{DF2}
f_{eq,||}(x,{\bf v}) = \frac{n_0 }{(2\pi)^{\frac{3}{2}} v_{th,p}^3} \exp\left\{-\frac{1}{2v_{th,p}^2}\left[v_x^2 + \left(v_y-U\left(x - 
\frac{v_z}{\Omega_{cp}}\right)\right)^2 + v_z^2 \right]\right\}
\end{equation}
This solution is a stationary equilibrium for the HVM set of equations Eqs. (\ref{vlasoveq}--\ref{E}), for the case of a shear 
parallel to the background magnetic field. Indeed Eqs. (\ref{FarAmp}--\ref{E}) are also exactly satisfied, since the $0$-th order moment 
of $f_{eq,||}$ gives a homogeneous density $n_0$ and the bulk speed ${\bf u}$ is along $y$ (see the next subsection for further details). 
The DF $f_{eq,||}$ is a Maxwellian-like function shifted along $v_y$, in which however the amount of the shift depends on both 
the position $x$ and the velocity $v_z$ (through the argument of $U$, $k_2=x -v_z/\Omega_{cp}$). For the same reason, the bulk speed $u_y$ 
does not coincide with $U$ in the general case. 

\subsection{Properties of the stationary distribution function $f_{eq,||}$}
To characterize the physical properties of $f_{eq,||}$, we focus here on its shape in velocity space and on the evaluation of its moments, 
in particular density, bulk velocity, temperature and heat flux.

\begin{figure*}[!t]
\epsfxsize=17cm \centerline{\epsffile{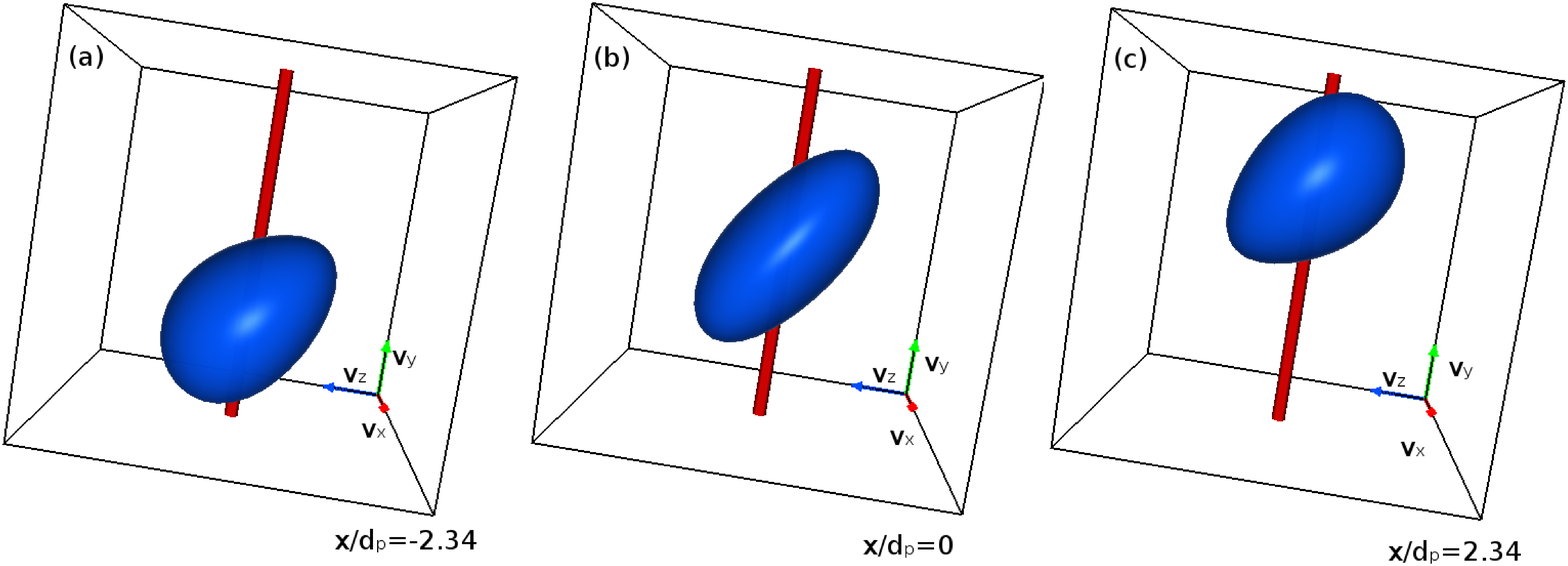}}   
\caption{Iso-surface plot of the proton DF $f_{eq,||}$ in velocity space at $x/d_p=-2.34$ (a), $x/d_p=0$ (b) and $x/d_p=2.34$ (c). In each 
panel, red, green and blue arrows refer to $v_x$, $v_y$ and $v_z$, respectively. The red tube indicates the magnetic field direction.}
\label{Fig:VDF}
\end{figure*}

To display the shape of $f_{eq,||}$ in the velocity space, we choose the $\tanh$-like shear profile of $U$, routinely adopted for 
investigating the KHI instability:
\begin{equation}
 U\left(k_2\right) = U_0\tanh{\left(\frac{k_2}{\Delta x}\right)}
 \label{shear0}
\end{equation}
where we choose $U_0=2V_{_A}$, $\beta_p=2v_{th,p}^2/V_{_A}^2=4$ and $\Delta x$ is the width of the shear function $U$. The width of the 
sheared bulk-velocity ${\bf u}(x)$ can be different from $\Delta x$. By using the latter expression of $U$, we compute $f_{eq,||}$ in Eq. 
(\ref{DF2}), by discretizing the four dimensional phase space through $N_x=512$ grid points in the one-dimensional spatial domain 
($x\in[-L/2,L/2]$) and $N_v=141$ grid points in each velocity direction ($v_j\in[-v_{max},v_{max}]$, being $j=x,y,z$ and 
$v_{max}=7v_{th,p}$), while we chose $L=50 d_p$ and $\Delta x = 2.5 d_p$. Figure \ref{Fig:VDF} show the iso-surface plots of $f_{eq,||}$ in 
velocity space at several spatial positions across the shear: $x/d_p=-2.34$ (a), $x/d_p=0.0$ (b) and $x/d_p=2.34$ (c). The red tube in Fig. 
\ref{Fig:VDF} indicates the magnetic field direction. Far from the shear (not explicitly reported) the distribution function is a shifted 
Maxwellian, while across the shear it becomes significantly stressed, resembling potato-like shapes with non-null heat flux, temperature 
anisotropy and agyrotropy with respect to its principal axes.

We also calculate the moments of the DF $f_{eq,||}$. We already anticipated that the density associated with the DF given by Eq. (\ref{DF2}) 
is uniform, i.e. $n(x) = n_0$. The bulk velocity can be easily evaluated, starting from ${\bf u} ({\bf x}) = \int d^3{\bf v}\, {\bf 
v} f_{eq,||}({\bf x},{\bf v})/n_0$. It can be easily shown that $u_x$ and $u_z$ are null, while $u_y$ is:
\begin{equation}\label{ustatx3}
u_y(x)= \frac{1}{(2\pi)^{1/2} v_{th,p}} \int_{-\infty}^{\infty} U\left( x - \frac{v}{\Omega_{cp}} \right) 
\exp \left( -\frac{v^2}{2 v_{th,p}^2} \right)dv
\end{equation}
The bulk velocity $u_y(x)$ associated with $f_{eq,||}$ does not coincide with the function $U(x)$, rather it is the result of the 
convolution between $U$ and a Gaussian function. 

Despite the inverse process - i.e. the determination of $U(x)$ and of $f_{eq,||}$ for a given profile of the bulk velocity $u_y(x)$ 
- requires the inversion of the convolution in Eq. (\ref{ustatx3}), it is still possible to deduce some simple results. If $\Delta x$ is 
the characteristic spatial length of $U(x)$, then $U(x-v/\Omega_{cp})$ - considered as a function of $v$ - varies over the scale $\Delta v = 
\Omega_{cp} \Delta x$. The Gaussian factor inside the integral of Eq. (\ref{ustatx3}) represents a windowing function selecting a $v$ 
interval of width $\sim v_{th,p}$; we indicate such a windowing function by:
\begin{equation}\label{wgauss}
W_G(v)=\frac{1}{(2\pi)^{1/2} v_{th,p}} \exp \left( -\frac{v^2}{2 v_{th,p}^2} \right)
\end{equation}

\noindent a)
In the large scale limit $\Delta x \gg \rho_{Lp}$, i.e. $\Delta v \gg v_{th,p}$, the profile of the windowing function is relatively 
unimportant and $W_G(v)$ can be successfully approximated with the simpler square window $W_S(v)$, centered in $v=0$, with width and 
amplitude $\sqrt{2\pi}v_{th,p}$ and $1/\sqrt{2\pi} v_{th,p}$, respectively. $W_G(v)$ and $W_S(v)$ have the same value at $v=0$ 
and the same integral in the interval $( -\infty , +\infty )$. Within the approximation $W_G(v) \simeq W_S(v)$, we obtain:
\begin{equation}\label{runave}
u_y (x) \simeq {\bar U}(x) = \frac{1}{\sqrt{2\pi}\rho_{_{Lp}}} \int_{-\sqrt{\frac{\pi}{2}}\rho_{_{Lp}}}^{\sqrt{\frac{\pi}{2}}\rho_{_{Lp}}} 
U(x-\xi)d\xi
\end{equation}
which represents the running average of $U(x)$ performed over the interval $\lbrack x-\sqrt{\pi/2} \rho_{_{Lp}},x+\sqrt{\pi/2} 
\rho_{_{Lp}}\rbrack$. In the large scale limit, the bulk velocity $u_y(x)$ is approximately given by the function $U(x)$ smoothed over 
an interval of amplitude $\sqrt{2\pi}\rho_{_{Lp}}$ centered at the position $x$. This result can be easily understood by thinking that the 
protons gyromotion in the plane perpendicular to ${\bf B}$ mixes up the $v_y$ velocities of the single protons, thus smoothing the profile 
of $U(x)$ over a length scale which is of the order of the Larmor radius. 

\noindent b)
In the opposite small scale limit, , i.e. $\Delta x \ll \rho_{_{Lp}}$,  Eq. (\ref{ustatx3}) can be rewritten as follows:
\begin{equation}\label{ustatx4}
u_y(x)= \frac{1}{(2\pi)^{1/2}} \int_{-\infty}^{\infty} U(\rho_{_{Lp}} \varphi) \exp \left[ -\frac{1}{2} \left(\varphi 
-\frac{x}{\rho_{_{Lp}}}\right)^2 
\right]d\varphi
\end{equation}
where $\varphi=k_2/\rho_{_{Lp}}$ and $U(\rho_{_{Lp}}\varphi)$ as a function of $\varphi$ varies on a scale much smaller than unity. If 
$U(\rho_{_{Lp}}\varphi)$ describes a shear layer corresponding to a bulk velocity which varies in the range $[-U_0,U_0]$, it can be 
approximated with the Heavyside function $H(\varphi)$:
\begin{equation}\label{Heavy}
U(\rho_{_{Lp}}\varphi) \simeq U_0 H(\varphi)=
\begin{cases}
    U_0   & \quad \text{if } \varphi>0 \\
   -U_0  & \quad \text{if } \varphi<0  
  \end{cases}
\end{equation}
After some algebraic steps, Eq. (\ref{ustatx4}) reduces to
\begin{equation}\label{ustatx6}
u_y(x)=
\begin{cases}
    U_0   & \quad \text{if } x\gg \rho_{_{Lp}} \\
   -U_0  & \quad \text{if } x\ll -\rho_{_{Lp}}  
  \end{cases}
\;\;\; , \;\;\; {\rm for}\; \Delta x \ll \rho_{_{Lp}}
\end{equation}
where, for simplifying the integrals, we considered that, for $x\gg \rho_{_{Lp}}$ ($x\ll -\rho_{_{Lp}}$), the Gaussian is essentially 
located in the positive (negative) part of the $\varphi$ axis, where $H(\varphi)=1$ ($H(\varphi)=-1$). Hence, in spite of the small scale 
of variation of the function $U(x)$ ($\Delta x \ll \rho_{_{Lp}}$), the bulk velocity $u_y(x)$ varies on a scale comparable with the 
proton Larmor radius $\rho_{_{Lp}}$. It is not possible to construct, in the current configuration, shear layers with a width smaller than 
the proton Larmor radius. This is again due to the proton gyromotion which mixes up the parallel velocity of single particles on a 
transverse scale of the order of $\rho_{_{Lp}}$.

\begin{figure}[!t]
\epsfxsize=8.5cm \centerline{\epsffile{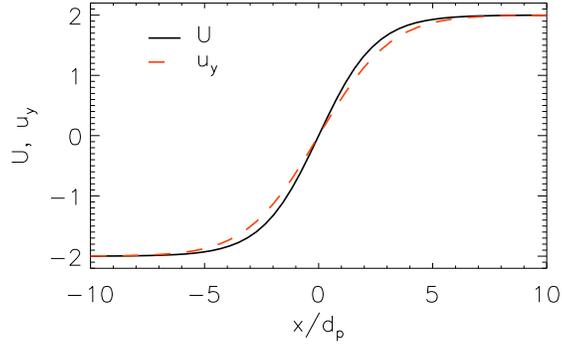}}   
\caption{Initial shear $U$ (black solid) and mean velocity $u_y$ (red dashed), evaluated by the proton DF $f_{eq,||}$, as a function of $x$ 
across the 
shear.}
\label{Fig:Uy}
\end{figure}

To directly display the shape of the bulk velocity, we consider the shear function $U(x)$ [Eq. (\ref{shear0})] and we numerically compute 
$u_y(x)$, for $\Delta x = 2.5 d_p  \simeq 1.77 \rho_{_{Lp}}$. Figure \ref{Fig:Uy} reports, the spatial profile of the function $U(x)$ 
(black solid line) and the corresponding bulk velocity $u_y$ (red dashed line). It is clear to note that significant differences between $U$ 
and $u_y$ are recovered. We also verified that in the large scale limit the windowing function does not play a significant role and hence, 
$u_y(x) \simeq U(x)$, while in the small scale limit protons arrange themselves to produce a bulk velocity $u_y(x)$ varying over a 
scale comparable with the proton Larmor radius, nevertheless the shear function $U$ varies over scales much smaller than $\rho_{_{Lp}}$ 
(not reported here).

To further characterize the moments of $f_{eq,||}$, we consider the variance matrix, defined by:
\begin{equation}\label{varmat}
\sigma_{ij}(x) = \frac{1}{n(x)} \int \left[ v_i - u_i(x) \right] 
\left[ v_j - u_j(x) \right] f_{eq,||}(x,{\bf v}) d^3 {\bf v} \;\; ; \;\; i,j = x,y,z 
\end{equation}
which is related to the proton temperature by $T_0 = m_p\sum_{j=1}^3 \sigma_{jj}/3k_B$. Since the magnetic field ${\bf B}$ is uniform and 
directed along $y$, the proton temperatures parallel and perpendicular to ${\bf B}$, i.e. in the local ${\bf B}$ frame (LBF), are defined, 
respectively, by $T_{||}=\frac{m_p}{k_B} \sigma_{yy}$ and $T_{\perp}=\frac{m_p}{k_B} (\sigma_{xx}+\sigma_{zz})/2$, so that $T_0 = 
(T_{||}+2T_{\perp})/3$.

The analytical evaluation of anisotropy and agyrotropy at the center of a symmetric shear, presented in App. \ref{app:anisev}, indicates 
that the equilibrium DF is anisotropic and agyrotropic. By means of the numerical evaluations of the variance matrix elements, we 
can extend the analytical computation and consider not only the center of the shear. We numerically diagonalize the matrix $\sigma$, thus 
rotating the DF into the minimum variance frame (MVF). The eigenvalues of $\sigma$, corresponding to the temperatures in the MVF, are 
indicated by: $\lambda^{(3)} < \lambda^{(2)} < \lambda^{(1)}$. 

\begin{figure}[!htb]
\epsfxsize=8.5cm \centerline{\epsffile{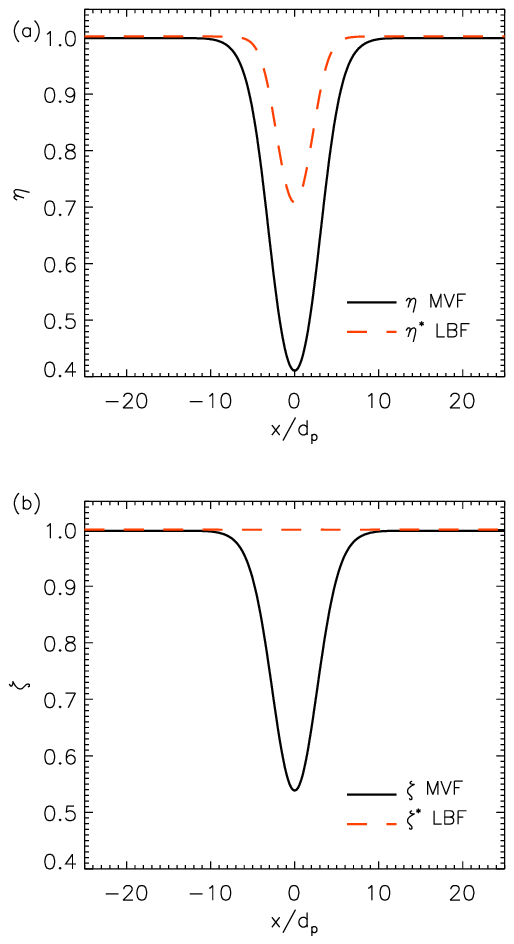}}   
\caption{Temperature anisotropy $\eta$, $\eta^*$ (top) and agyrotropy $\zeta$, $\zeta^*$ (bottom) evaluated in the minimum variance frame 
(black solid) and in the local magnetic field frame (red dashed), associated with the proton DF $f_{eq,||}$.}
\label{Fig:temp}
\end{figure}

The top panel of Fig. \ref{Fig:temp} shows the temperature anisotropy ratio $\eta$ and $\eta^*$, while the bottom panel indicates the 
agyrotropy ratio $\zeta$ and $\zeta^*$, for the shear $U(x)$ given by Eq. (\ref{shear0}) in the case $\Delta x=2.5d_p$, both in the MVF 
(black solid) and in the LBF (red dashed). Temperature anisotropy and agyrotropy have been evaluated as follows: (a) temperature anisotropy 
in the MVF $\eta=(\lambda_2+\lambda_3)/2\lambda_1$; (b) temperature anisotropy in the LBF $\eta^*=(\sigma_{xx}+\sigma_{zz})/2\sigma_{yy}$; 
(c) agyrotropy in the MVF $\zeta =\lambda_3/\lambda_2$; (d) agyrotropy in the LBF $\zeta^* 
=\min(\sigma_{xx},\sigma_{zz})/\max(\sigma_{xx},\sigma_{zz})$. In the LBF, the DF is anisotropic close to the shear, while no temperature 
agyrotropies are recovered. On the other hand, in the MVF, the DF displays strong anisotropies as well as agyrotropies close to the velocity 
shear.

\begin{figure}[!htb]
\epsfxsize=8.5cm \centerline{\epsffile{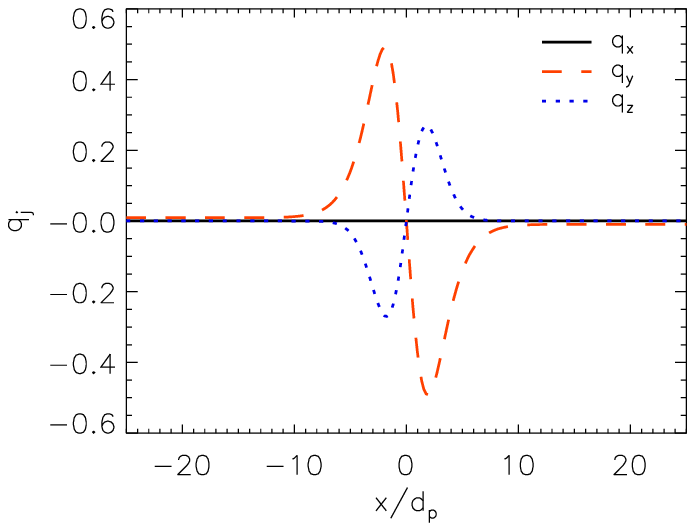}}   
\caption{Heat flux $q_x$ (black solid), $q_y$ (red dashed) and $q_z$ (blue dotted), associated with the proton DF $f_{eq,||}$.}
\label{Fig:heatfl}
\end{figure}

We finally characterize the DF $f_{eq,||}$ by computing the heat flux for unit of mass:
\begin{equation}\label{heatfl}
q_j(x) = \frac{1}{2} \int \left[v_j - u_j({x}] \right) \left[ {\bf v} - {\bf u}(x) \right]^2 f_{eq,||}(x,{\bf v}) d^3 {\bf v} \;\; ; \;\; 
j= x,y,z \end{equation}
where the shear function $U(x)$ in Eq. (\ref{shear0}), with $\Delta x= 2.5 d_p$, is adopted. Figure \ref{Fig:heatfl} reports the three 
components of the heat flux $q_x$ (black-solid line), $q_y$ (red-dashed line) and $q_z$ (blue-dotted line), as a function of $x/d_p$. The 
equilibrium DF $f_{eq,||}$ is such that a non-vanishing heat flux is recovered at $x/d_p\simeq 0$ in the two components $q_y$ and $q_z$, 
which tends to zero away from the shear.

\section{Stationary solution for the perpendicular case: ${\bf u_0}\, \perp \, {\bf B}$}
\label{sect:perpEQDF}

We revisit here the derivation of the stationary solution of Eqs. (\ref{vlasoveq}--\ref{E}), in the case in which velocity shear and 
homogeneous magnetic field are perpendicular, while the electric field is ${\bf E} = E(x) {\bf e}_x$. The method here adopted is based 
on the dynamics of a single proton in the electric ${\bf E}$ and magnetic ${\bf B}$ field. Particle trajectories have been already studied 
to build up a stationary solution in a fully kinetic (ion+electron) description in previous studies \cite{ganguli88,nishikawa88,cai90}. 
In particular, in the analytical description of single particle dynamics we follow the same method as Ganguli et al. \cite{ganguli88}, but 
deriving further general results which are important for setting up our numerical description of particle dynamics. We will also derive a 
form for the proton distribution function which is different from that in Ref. \cite{ganguli88} (except in the particular case of linearly 
growing electric field). In our case the derived DF is furthermore supplemented with a form for the electron pressure profile, which allows 
us to obtain an exact stationary state for the whole set of HVM equations.

Our derivation starts from considering the single-particle motion. The proton motion along the parallel $z$ direction is decoupled from 
the motion in the transverse plane. Therefore, we focus on the particle motion in the $xy$ plane, described by:
\begin{eqnarray}\label{motioneq}
\frac{dv_x}{dt} &=& \Omega_{cp} v_y + \frac{e}{m_p} E(x) \nonumber \\
\frac{dv_y}{dt} &=& -\Omega_{cp} v_x \\
\frac{dx}{dt} &=& v_x \nonumber
\end{eqnarray}
The particle motion depends on the electric field profile $E(x)$, which indirectly determines the profile $u(x)$ of the bulk velocity. We 
consider a situation where $u(x)$ varies crossing one or more shear layers, but becomes essentially uniform far from the shear layers: 
$E(x)= E_{+\infty}$ ($E(x)=E_{-\infty}$) for large positive (negative) $x$. Thus, in the homogeneous region particles drift along $y$ with a 
uniform drift velocity ${\bf v}_{d,\pm\infty} = c({\bf E_{\pm\infty}}\times {\bf B})/B^2$.

From Eqs. (\ref{motioneq}), it is easy to obtain:
\begin{eqnarray}
v_y &=& -\Omega_{cp} x + W_0  \label{constmot} \\
\frac{d^2 x}{dt^2} &=& -\Omega_{cp}^2 x + \frac{e}{m_p} E(x) + \Omega_{cp} W_0 \label{eqx} 
\end{eqnarray}
where $W_0$ is a constant determined by initial conditions and Eq. (\ref{eqx}) is a nonlinear oscillator equation for $x(t)$. We integrate 
Eq. (\ref{eqx}) in the interval $\left[ x_0,x\right]$, by considering that $dv_x/dt = d/dx(v_x^2/2)$ and by rewriting 
$v_{0y}=W_0-\Omega_{cp} x_0$ (being $x_0$ an arbitrary position corresponding to $v_{0y}$):
\begin{equation}\label{cons3} 
\frac{1}{2} m_p v_x^2 + e \phi(x) + \frac{1}{2} m_p \Omega_{cp}^2 (x-x_0)^2 -m_p\Omega_{cp} v_{0y} (x-x_0) = e_0
\end{equation}
where $\phi(x)= -\int_{x_0}^x E(x') dx'$ is the electrostatic potential, which vanishes at $x=x_0$, and $e_0=m_p v_{x0}^2/2$ is a constant, 
$v_{x0}$ being the value of $v_x$ at $x=x_0$. 

Eq. (\ref{cons3}) expresses the energy conservation for a particle with mass $m_p$ following a 1D motion in the effective potential energy 
$U_{\rm eff}(x) = U_E(x) + U_\Omega(x)$, where $U_E(x)=e\phi(x)$ is the electrostatic potential energy and 
\begin{equation}\label{UOmega}
U_\Omega(x) = \frac{1}{2} m_p \Omega_{cp}^2 (x-x_0)^2 -m_p\Omega_{cp} v_{0y} (x-x_0) =
\frac{1}{2} m_p \Omega_{cp}^2 \left( x- \frac{W_0}{\Omega_{cp}} \right)^2 + \frac{1}{2} m_p v_{0y}^2
\end{equation}
The term $U_\Omega(x)$ hence corresponds to an attractive force towards the position $W_0/\Omega_{cp}$.

For large $|x-x_0|$ the term $U_\Omega(x)$ dominates in the determining the effective potential energy $U_{\rm eff}(x)$. Indeed, since 
$E(x)$ becomes constant for sufficiently large values of $|x-x_0|$, we have that $U_E \simeq -E_{+\infty}(x-x_0)$ [$U_E \simeq 
-E_{+\infty}(x-x_0)$] for large positive (negative) values of $x-x_0$, while $U_\Omega(x)$ is quadratic with respect to $(x-x_0)$. The 
motion of the particle along $x$ is hence confined inside a potential well: $x_m \le x(t) \le x_M$, where $U_{\rm eff}(x_{m})=U_{\rm 
eff}(x_{M})=e_0$ and the particle moves back and forth in the interval $\left[ x_m , x_M \right]$, with vanishing $v_x$ at $x_m$ and $x_M$. 
In other words, $x(t)$ and $v_x(t)$ are periodic function with period $\tau$ and, from Eq. (\ref{constmot}), follows that $v_y(t)$ is also 
periodic with period $\tau$. Therefore, the particle follows a closed trajectory in the $v_xv_y$ plane. Notice that $y(t)$ is not 
necessarily a periodic function and the particle trajectory in the $xy$ plane is, in general, an open curve. The details of the motion along 
$x$, like the period $\tau$, depend both on the specific form of the electric field profile $E(x)$ and on the particle initial conditions, 
which determine the constant quantity $W_0$. However, the periodicity of variables $x(t)$, $v_x(t)$ and $v_y(t)$ holds for any form of 
$E(x)$ and for any initial condition. This result is crucial for the setup of the numerical method we employed to calculate a stationary 
proton DF $f_{eq,\perp}$ for an arbitrary electric field profile.

Since the particle motion in the $v_xv_y$ plane is periodic, the time average of the velocity over the period $\tau$ provides the drift 
velocity in the particle motion. Therefore, we define the guiding center velocity ${\bf v}_c$ as: 
\begin{equation}\label{vc}
{\bf v}_c = \langle {\bf v} \rangle_\tau = \frac{1}{\tau} \int_0^\tau {\bf v}(t)\, dt
\end{equation}
The $x$-component of the guiding center velocity is trivially vanishing, hence: ${\bf v}_c=v_{cy} {\bf e}_y = \langle v_y \rangle_\tau\, 
{\bf 
e}_y$. We also define the {\em guiding center position} $x_c$ as the position where the particle velocity component $v_y$ is equal to the 
guiding center velocity: $v_y=v_{cy}$; hence, from Eq. (\ref{constmot}),  we find
\begin{equation}\label{xc}
x_c = \left(W_0-v_{cy}\right)/\Omega_{cp}
\end{equation}
Note that: (i) equation (\ref{xc}) implies that a single value is admitted for $x_c$; (ii) the guiding center position $x_c$ 
represents also the time-averaged particle $x$-position: $x_c=\langle x \rangle_\tau$.

We point out that our particle guiding center definition is different from that used in previous studies. In fact, in Refs. \cite{naitou80,cai90} the 
guiding center position is defined as a point where $v_y=u_y$, implying that a given particle can have more than one guiding center (see 
Ref. \cite{cai90} for a discussion). In contrast, in our approach a single guiding center is defined for each particle, regardless of 
the specific electric field profile $E(x)$ and of the particle initial condition. In Ref. \cite{ganguli88} the guiding center position is 
defined as the position where the effective potential energy is minimum. Also this definition differs from ours, except for particular 
profiles of the linearly growing electric field (see App. \ref{app:LAeq}).

To build the stationary DF for the HVM Eqs. (\ref{vlasoveq})-(\ref{E}), we consider the particle total energy, which is a 
constant of motion:
\begin{equation}\label{toten}
\mathscr{E}=K+U'_E = \frac{1}{2}m_p \left( v_x^2 + v_y^2 + v_z^2 \right) + U'_E
\end{equation}
where the electric potential energy $U'_E(x)$ is re-defined such that $U'_E(x_c)=0$, i.e. it vanishes at the guiding center position $x_c$ 
of the considered particle $U'_E(x)=-e \int_{x_c}^x E(x')\, dx'$. This choice can be justified by the following argument. Let us consider 
the particular case of a uniform electric field $E(x)=E_0$, corresponding to proton circular orbits in the $v_xv_y$ plane, with uniform 
drift velocity $v_{cy}=-cE_0/B$. In such a case, the potential energy has the form $U'_E(x)=-e E_0 (x-x_c)$. Since $x_c=\langle x 
\rangle_\tau$, it follows that $\langle U'_E \rangle_\tau=0$, i.e. it has the same value for all particles, regardless of their average 
position $x_c$. This is in accordance with the macroscopic invariance of fluid properties with $x$, which characterizes this particular 
case. In contrast, a potential energy $U_E$ which vanishes at a fixed position $x_0$ (equal for all the particles) would give $\langle U_E 
\rangle_\tau=eE_0(x_c-x_0)$, i.e., an average potential energy which systematically varies with the average position $x_c$ of particles.

We define also the quantity $\mathscr{E}_{0}=\mathscr{E}-m_p v_{cy}^2/2$ which represents the part of the particle energy not due to the 
drifting motion. 
Of course, $\mathscr{E}_0$ is another motion constant. Since $\mathscr{E}$ is constant, its value can be evaluated at $x=x_c$, where 
$U'_E(x_c)=0$ and 
where, by definition, $v_y=v_{cy}$. Thus:
\begin{equation}\label{toten2}
\mathscr{E}=\frac{1}{2}m_p \left\{ \left[v_x(x=x_c)\right]^2 + v_z^2 \right\} + \frac{1}{2}m_p v_{cy}^2 
\end{equation}
implying 
\begin{equation}\label{e}
\mathscr{E}_0=\frac{1}{2}m_p \left\{ \left[v_x(x=x_c)\right]^2 + v_z^2 \right\}
\end{equation}

To understand how to define the proton DF, we again consider the particular case of constant electric field and we require that, in this 
case, the DF is a shifted Maxwellian:
\begin{equation}\label{shiftMax}
f_{SM}({\bf v}) = C \exp \left[ -\frac{v_x^2+(v_y-v_{cy})^2+v_z^2}{2 v_{th,p}^2}\right]
\end{equation}
where $v_{th,p}$ is the thermal speed and $C$ is a normalization constant. The uniform-$E$ case above analyzed can be also reproduced in 
the local approximation description, whose details are reported in App. \ref{app:LAeq}, by setting $\alpha_0=0$ 
($\omega^2=\Omega_{cp}^2$), which implies that $v_x^2+(v_y-v_{cy})^2=\left[ v_x(x=x_c)\right]^2$. Therefore, in this case, by comparing 
Eq. (\ref{e}) and Eq. (\ref{shiftMax}), we conclude that:
\begin{equation}\label{shiftMax2}
f_{SM}({\bf v}) = C \exp \left( -\frac{\mathscr{E}_0}{m_p v_{th,p}^2}\right)
\end{equation}
The above considerations lead us to the following ``ansatz": we hypothesize that a stationary proton DF representing a shearing flow for 
{\em any} electric field profile $E(x)$ can have the following implicit form:
\begin{equation}\label{fgen}
f_{eq,\perp}(x,{\bf v})=C \exp \left[ -\frac{\mathscr{E}_0(x,v_x,v_y,v_z)}{m_p v_{th,p}^2}\right]
\end{equation}
where $\mathscr{E}_0(x,v_x,v_y,v_z)=\mathscr{E}-m_p v_{cy}^2/2$ is the single-proton energy not due to the drifting motion. As showed above, 
Eq. (\ref{fgen}) is a shifted Maxwellian for an uniform electric field. 

Of course, such a conjecture must be verified {\em a posteriori}. This can be done first in the particular case of the local approximation, 
reported in App. \ref{app:LAeq}, where we also derive the explicit form of the equilibrium DF. Then, in the case of a 
generic electric field profile $E(x)$ the same quantities will be calculated by employing a numerical technique. It is important to 
highlight that, since the quantity $\mathscr{E}_0$ is a constant of the particle motion, $f_{eq,\perp}$ is a stationary solution of the 
Vlasov equation [Eq. (\ref{vlasoveq})], provided that both the electric and magnetic fields are temporally constant \cite{krall73}.

The numerical method employed to generate the equilibrium DF $f_{eq,\perp}$, for a generic electric field profile $E(x)$, is described 
in the following. We assume that $f_{eq,\perp}$ has the form given by Eq. (\ref{fgen}), being 
\begin{equation}\label{enum}
\mathscr{E}_0(x,v_x,v_y,v_z)=\frac{m_p}{2} \left( v_x^2+v_y^2+v_z^2 \right) +U'_E(x,v_x,v_y)-\frac{m_p}{2} v_{cy}^2(x,v_x,v_y)
\end{equation}
Here, $v_{cy}^2(x,v_x,v_y)$ indicates the guiding center velocity of a particle which is located at the position $x$, with velocity 
$(v_x,v_y)$ at a given time $t$, while $U'_E(x,v_x,v_y)$ is the electric potential of the same particle. Since $\mathscr{E}_0$ is a motion 
constant, the time $t$ when $\mathscr{E}_0$ is evaluated can be arbitrarily chosen. Hence, in Eq. (\ref{enum}), $(x,v_x,v_y,v_z)$ can be 
interpreted as the position and velocity of a single particle at the initial time of its motion. The evaluation of the last two terms in 
Eq. (\ref{enum}) requires however to integrate the single particle motion Eqs. (\ref{motioneq}). This has been done by the following 
numerical procedure:

\noindent
(i) We consider a 1D-3V phase space, composed by a spatial coordinate $x\in[0,L]$, discretized with $N_x$ grid points, and three 
velocity coordinates $v_m \in \left[ -v_{max},v_{max} \right] \, m=x,y,z$, discretized with $N_v$ grid points along each direction. We 
numerically integrate $N_x\times N_v^2$ particle motion Eqs. (\ref{motioneq}), using each point $(x_i,v_{x,j},v_{y,k})$ of the subgrid as 
initial condition: $x(t=0)=x_i$, $v_x(t=0)=v_{x,j}$, $v_y(t=0)=v_{y,k}$ ($i,j,k$ are indexes which span along $x$, $v_x$ and $v_y$, 
respectively). The $v_z$ component, whose index is $l$; is neglected since the motion is trivial along $z$. The time integration of Eqs. 
(\ref{motioneq}) has been carried out through a $3$-order Adam-Bashforth scheme, being the time step $\Delta t$ chosen to maintain the CFL 
condition.

\noindent
(ii) Since the exact trajectories in the $v_x v_y$ plane are necessarily closed, each integration is stopped when the corresponding 
orbit in the $v_x v_y$ plane is completed. The corresponding time $T$ represents the orbit period. 

\noindent
(iii) We calculate, for each orbit, the velocity and $x$-position of the guiding center: $v_{cy,ijk}=\langle v_y \rangle_T$ and 
$x_{c,ijk}=\langle x \rangle_T$. The electric potential associated with the particle initial position is calculated as 
$\phi_{ijk}=-\int_{x_{c,ijk}}^{x_i} E(x)\, dx$, where the integral is numerically evaluated. Then, the DF at a given point of the phase 
space is evaluated as
\begin{equation}\label{fnum}
f_{eq,\perp}(x_i,v_{x,j},v_{y,k},v_{z,l}) = C \exp \left[ -\frac{1}{2 v_{th,p}^2} \left( v_{x,j}^2 + v_{y,k}^2 + v_{z,l}^2 - v_{yc,ijk}^2 
\right) 
-\frac{e \phi_{ijk}}{m_p v_{th,p}^2} \right]
\end{equation}
From this expression, the moments of the distribution function (density, temperatures, bulk velocity and heat flux components) are 
numerically evaluated. In particular, all the moments vary only in the $x$ direction and the bulk velocity ${\bf u}(x)$ is directed in the 
$y$ direction.

\noindent
(iv) The resulting $u_y(x)$ depends on the chosen profile of the electric field $E(x)$. However, for an arbitrary electric field profile 
$E(x)$, the bulk velocity does not coincides with the local ${\bf E}\times {\bf B}$ drift velocity. On the other hand, 
${\bf E}$ and ${\bf u}$ appear in the generalized Ohm's law (\ref{E}), which must be consistently satisfied. This is obtained by 
choosing a profile for the electron pressure $p_e(x)$ such that
\begin{equation}\label{peprof}
\frac{dp_e}{dx}=-qn(x) \left[ \frac{Bu_y(x)}{c}+E(x) \right]
\end{equation}
Then, in the general case the electron pressure $p_e$ is not uniform, except in the case of a linear electric field profile (See App. 
\ref{app:LAeq}). By adopting this closure for the electron pressure, it is easy to show that the considered configuration is a stationary 
solution of the whole set of HVM Eqs. (\ref{vlasoveq})-(\ref{E}) supplemented by Eq. (\ref{eadiab}) for the electron pressure.

\subsection{Properties of the stationary distribution function $f_{eq,\perp}$}

To analyze the properties of the equilibrium DF $f_{eq,\perp}$, we focus on the following shape for the electric field: 
\begin{equation}\label{Enum0}
 E_{x}(x) = - E_0 \tanh{\left( \frac{x}{\Delta x} \right)} 
\end{equation}
representing a shear layers of amplitude $E_0=1$ (in scaled unit) and width $\Delta x=2.5 d_p$. By using the latter expression of $E_{0,x}$, 
we compute $f_{eq,\perp}$ in Eq. (\ref{fgen}), by discretizing the four dimensional phase space through $N_x=512$ grid points in the 
one-dimensional spatial domain ($x\in[-L/2,L/2]$) and $N_v=141$ grid points in each velocity direction ($v_j\in[-v_{max},v_{max}]$, being 
$j=x,y,z$ and $v_{max}=7v_{th,p}$), while we chose $L=50 d_p$ and $\Delta x = 2.5 d_p$. It is worth to note that the difference between 
the electric field from which we compute the equilibrium [Eq. (\ref{Enum0})] and the term  $-{\bf u} \times {\bf B}/c$ evaluated using the 
equilibrium DF mean speed is of the order of $10^{-5}$. Despite this quantity is small at the initial time, one needs to take 
care of it by self-consistently solving Eq. (\ref{eadiab}) to maintain the equilibrium.

\begin{figure*}[!t]
\epsfxsize=17cm \centerline{\epsffile{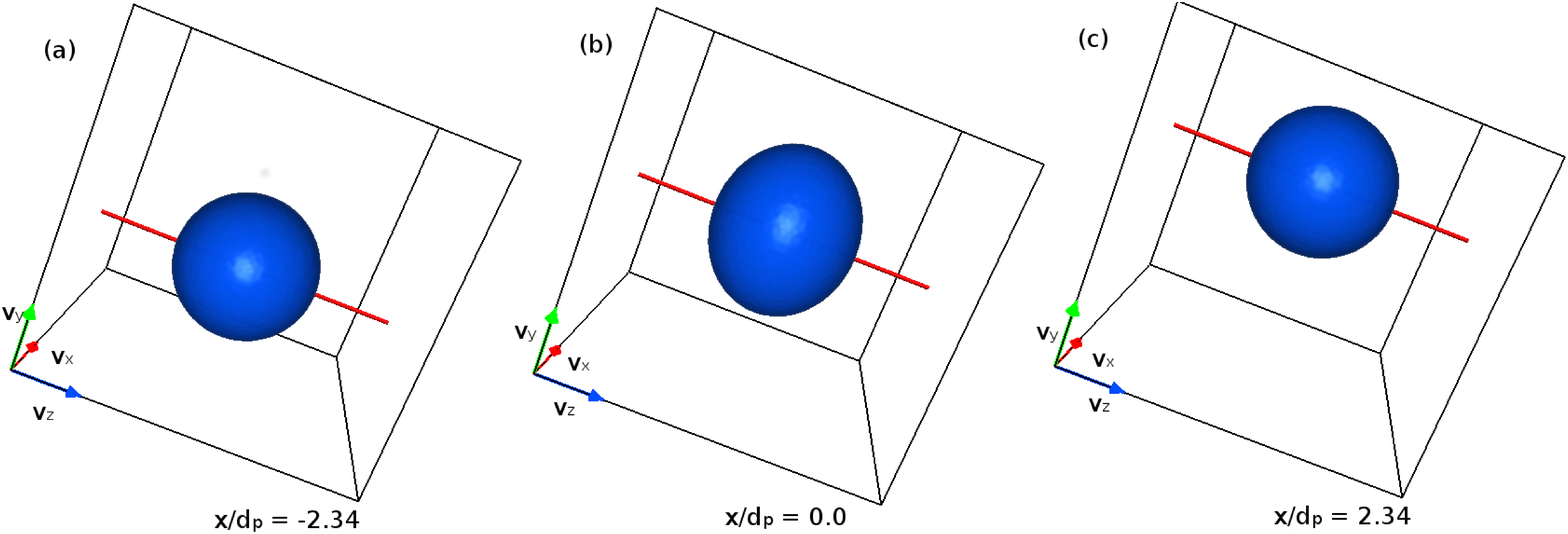}}   
\caption{Iso-surface plot of the proton DF $f_{eq,\perp}$ in velocity space at $x/d_p=-2.34$ (a), $x/d_p=0$ (b) and $x/d_p=2.34$ (c). In 
each panel, red, green and blue arrows refer to $v_x$, $v_y$ and $v_z$, respectively. The red tube indicates the magnetic field direction.}
\label{Fig:VDFBperpU}
\end{figure*}

Figure \ref{Fig:VDFBperpU} reports iso-contour of the proton DF $f_{eq,\perp}$ in the velocity space. Panels (a) to (c) refer to 
different positions across the shear: $x/d_p=-2.34$ (a), $x/d_p=0.0$ (b) and $x/d_p=2.34$ (c). The red tube in Fig. \ref{Fig:VDFBperpU} 
indicates the magnetic field direction. We note that, against the parallel case, here the equilibrium DF is less stressed and exhibit a 
bi-Maxwellian like structure, elongated in a direction transverse to the magnetic field direction. As for the parallel case, far from the 
shear, the DF $f_{eq,\perp}$ reduces to the shifted Maxwellian, while - across the shear - it exhibits a clear temperature anisotropy.

\begin{figure}[!t]
\epsfxsize=8.5cm \centerline{\epsffile{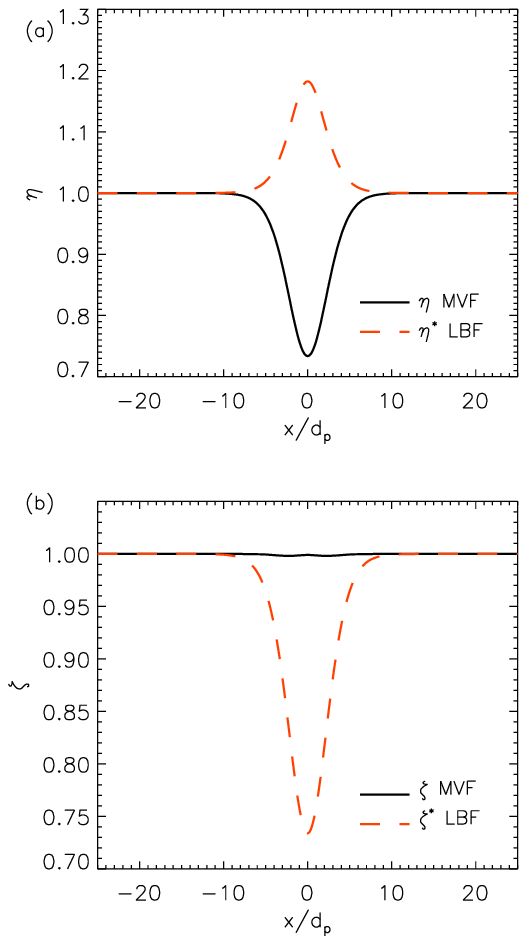}}   
\caption{Temperature anisotropy $\eta$, $\eta^*$ (top) and agyrotropy $\zeta$, $\zeta^*$ (bottom) evaluated in the minimum variance 
frame (black solid) and in the local magnetic field frame (red dashed), associated with the proton DF $f_{eq,\perp}$.}
\label{Fig:tempBperpU}
\end{figure}

Figure \ref{Fig:tempBperpU} reports the temperature anisotropy (top panel) and agyrotropy (bottom panel) ratios both in the MVF (black 
solid) and in the LBF (red dashed). Temperature anisotropy and agyrotropy have been evaluated as follows: (a) temperature anisotropy in the 
MVF $\eta=(\lambda_2+\lambda_3)/2\lambda_1$; (b) temperature anisotropy in the LBF $\eta^*=(\sigma_{xx}+\sigma_{yy})/2\sigma_{zz}$; (c) 
agyrotropy in the MVF $\zeta =\lambda_3/\lambda_2$; (d) agyrotropy in the LBF $\zeta^* 
=\min(\sigma_{xx},\sigma_{yy})/\max(\sigma_{xx},\sigma_{yy})$. Note that the definitions in the LBF are different from the parallel case, 
since the orientation of the magnetic field has been changed. If in the parallel case the equilibrium distribution function was 
characterized by stronger anisotropies in the MVF frame, here the situation is opposite. Indeed, in the MVF, the DF is strongly anisotropic 
at the shear, but it does not present significant non-gyrotropic features. On the other hand, in the LBF frame, the DF is significantly 
anisotropic as well agyrotropic.

\begin{figure}[!b]
\epsfxsize=8.5cm \centerline{\epsffile{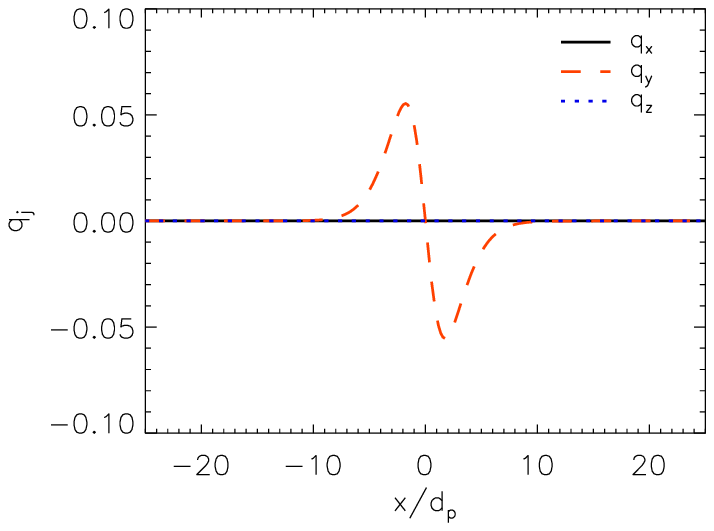}}   
\caption{Heat flux $q_x$ (black solid), $q_y$ (red dashed) and $q_z$ (blue dotted), associated with the proton DF $f_{eq,\perp}$.}
\label{Fig:heatflBperpU}
\end{figure}

Finally, we characterize the DF by evaluating the heat flux [Eq. (\ref{heatfl})]. Figure \ref{Fig:heatflBperpU} reports the three components of the 
heat flux $q_x$ (black-solid line), $q_y$ (red-dashed line) and $q_z$ (blue-dotted line), as a function of $x/d_p$. Clearly, a non-vanishing 
heat flux is recovered at $x/d_p \simeq 0$ in the $y$ component.

\section{Hybrid Vlasov-Maxwell simulations of the equilibrium}

In this section we numerically test that distribution functions derived in the previous sections $f_{eq,||}$ and $f_{eq,\perp}$ are 
effectively stationary equilibria for the HVM set of equation, which are solved numerically in the so-called HVM code \cite{valentini07, 
cerri17,pezzi17b,pezzi17c}, by also comparing these equilibria with the sheared Maxwellian case $f_{SM}$. 

The HVM code solves numerically the set of Eqs. (\ref{vlasoveq}--\ref{eadiab}) in dimensionless form through a Eulerian algorithm described 
in detail in Ref. \cite{valentini07}. In the parallel case, since $T_e$ is homogeneous and constant, Eq. (\ref{eadiab}) is trivial. 
Dimensionless HVM equations are obtained by scaling velocities by the Alfv\'en speed $V_{_A}$, lengths by the proton skin depth $d_p$ and 
time by the inverse proton cyclotron frequency $\Omega_{cp}^{-1}$. Since the problem is intrinsically one-dimensional in physical space, we 
restrict our numerical runs to a phase space of reduced dimensionality (1D in physical space and 3D in velocity space). The code assumes 
periodic boundary conditions in the spatial coordinate $x\in[0,L]$, while the DF $f(x,{\bf v},t)$ is set equal to zero for $|v|>v_{max}$ in 
each velocity direction and at each spatial position. 

\subsection{Parallel case}
For this case, we discretized the four dimensional phase space through $N_x=512$ grid points in the one-dimensional spatial domain and 
$N_v=141$ grid points in each velocity direction, while $v_{max}=7 v_{th,p}$. 

We performed two simulations (S1 and S2) keeping fixed the background magnetic field ${\bf B}_0=B_0 {\bf e}_y$ ($B_0=1$ in scaled units) and 
the proton plasma beta $\beta_p=2v_{th,p}^2/V_{_A}^2=4$, but changing the initial proton distribution function. In both simulations the 
system dynamics is followed up to a time $t_{max}=40\Omega_{cp}^{-1}$ and no perturbations are introduced.

We first consider the SMDF $f_{SM}$ [Eq. (\ref{PDFatt})] as initial condition for the simulation S1. In this case we set: 
\begin{equation}
 u_0(x)=U_0\left[\tanh{\left(\frac{x-L/4}{\Delta x}\right)}-\tanh{\left(\frac{x-3L/4}{\Delta x}\right)} -1 \right]
 \label{shear1}
\end{equation}
with $U_0=2V_{_A}$, $\Delta x=2.5 d_p$ and $L=100 d_p$.

Then, we performed a second simulation S2, using as initial condition the stationary DF $f_{eq,||}$ [Eq. (\ref{DF2})], with
\begin{equation}
 U(x-v_z)=U_0\left[ \tanh{\left(\frac{x-L/4-v_z}{\Delta x}\right)}-\tanh{\left(\frac{x-3L/4-v_z}{\Delta x}\right)} - 1\right]
 \label{shear2}
\end{equation}
and compared the results of the two simulations. We point out that the expressions in Eqs. (\ref{shear1}) and (\ref{shear2}) 
describe a smooth jump in velocity at the position $x=L/4$ along the $x$-direction; this jump has been mirrored at $x=3L/4$, in 
order to satisfy the periodic boundary conditions.

\begin{figure*}[!htb]
\epsfxsize=12.5cm \centerline{\epsffile{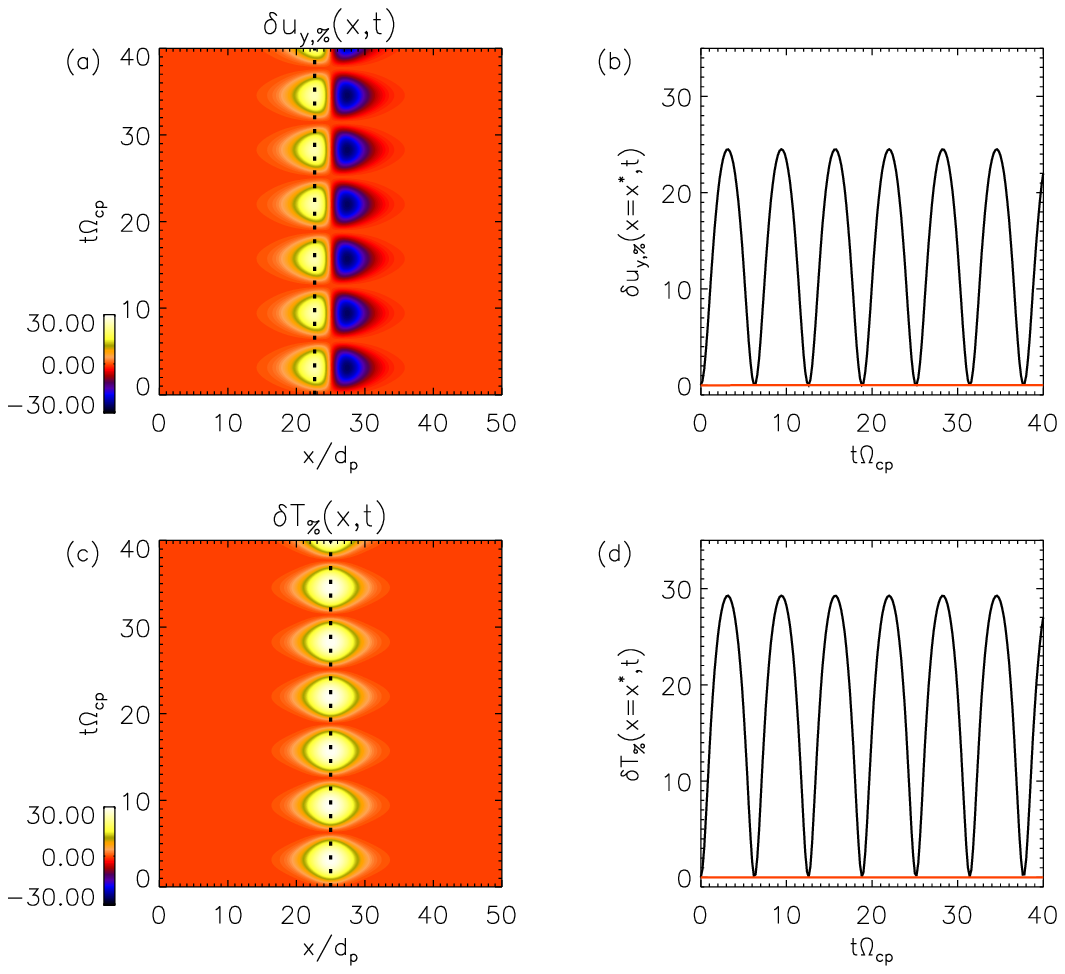}}   
\caption{Top: contour plot of $\delta u_{y,\%}(x,t)$ (a) and the temporal profile of $\delta u_{y,\%}(x=x^*,t)$ (b), being $x=x^*$ 
indicated in panel (a) with the black dashed line. Bottom: contour plot of $\delta T_{\%}(x,t)$ (c) and the temporal profile of $\delta 
T_{\%}(x=x^*,t)$ (d), being $x=x^*$ indicated in panel (c) with the black dashed line.}
\label{Fig:eq}
\end{figure*}

As expected, the simulation S1 clearly indicates that the initial distribution SM DF is not an equilibrium and, as a consequence, its 
velocity moments display an oscillatory behaviour with a period equal to the ion gyroperiod. Left column of Fig. \ref{Fig:eq} displays the 
contour plots of $\delta u_{y,\%}(x,t) = (u_y(x,t)-u_y(x,0))/U_0\times100$ [panel (a)] and $\delta 
T_{\%}(x,t)=(T(x,t)-T(x,0))/T^{\infty}\times100$ [panel (c)] in the $(x,t)$ plane, for the simulation S1, where the $x$ range has been 
set to focus on the left half of the spatial box and where $T^{\infty} = v_{th,p}^2= \beta_p/2 = 2$ in dimensionless units. Significant 
oscillations (about $30\%$) are recovered in both $u_y$ and $T$, localized around the shear.

On the other hand, in the simulations S2 with the DF $f_{eq,||}$, the system remains at equilibrium (no significant oscillations are 
visible). To better point out the differences between S1 and S2, we considered the temporal profiles of $\delta u_{y,\%}(x,t)$ and $\delta 
T_{\%}(x,t)$, evaluated at a fixed spatial position $x=x^*$. The point $x^*$ corresponds to the spatial point where each quantity 
exhibits the largest amplitude oscillations in simulation S1 (vertical black-dashed lines in panels (a) and (c) of Fig. \ref{Fig:eq}), with 
respect to the initial condition, i.e. $\delta u_{y,\%}(x^*,t)=\max_x\{\delta u_{y,\%}(x,t)\}$ and $\delta T_{\%}(x^*,t)=\max_x\{\delta 
T_{\%}(x,t)\}$. These temporal profiles are reported in panels (b) and (d) of Figure \ref{Fig:eq}, as black ($f_{SM}$) and red 
($f_{eq,||}$) curves, respectively. Here, one can realize that no significant oscillations in the signals are recovered in the case of the 
DF $f_{eq,||}$, as compared to the case of the SMDF.

\begin{figure}[!htb]
\epsfxsize=7cm \centerline{\epsffile{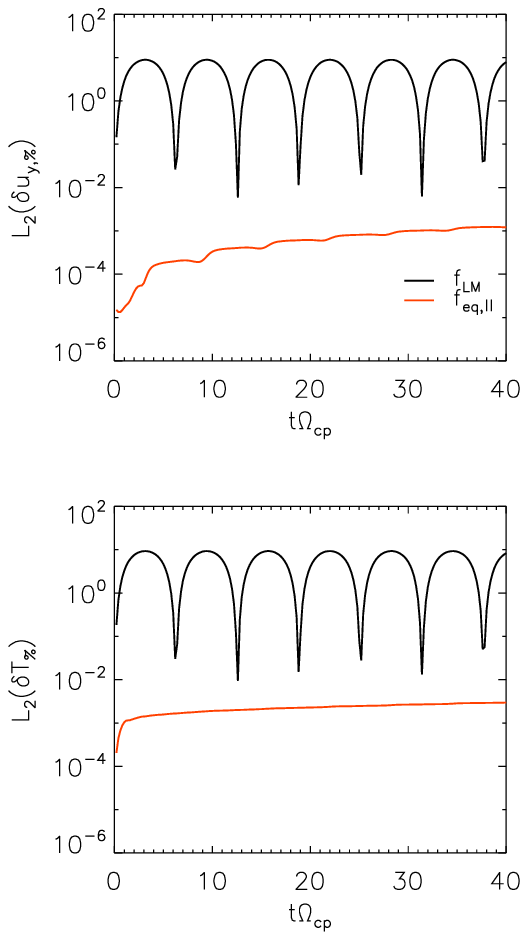}}   
\caption{Temporal evolution of $L_2(\delta u_{y,\%})$ (top row) and $L_2(\delta T_{\%})$ (bottom row). In each panel the black line 
refer to the $f_{SM}$ DF, while the red line to the $f_{eq,||}$ DF.}
\label{Fig:norme}
\end{figure}

To further characterize the differences between the two cases S1 and S2, we also computed the $L_2$ norm of $\delta u_{y,\%}(x,t)$ and 
$\delta T_{\%}(x,t)$, defined as $L_2(g(x,t)) = \sqrt{\int (g-g_0)^2 dx / L} $, being $g$ a generic function and $g_0=g(x,0)$. 
Clearly $L_2(g(x,t))$ is a function of time $t$. In Figure \ref{Fig:norme}, we show, in semi-logarithmic plot, the temporal evolution of 
$L_2(\delta u_{y,\%})$ (top row) and $L_2(\delta T_{\%})$ (bottom row). As it can be appreciated from the plots in this figure, significant 
oscillations with respect to the initial configuration are present in the case of the SMDF (black-solid curves), confirming that this 
distribution is not a HVM equilibrium. On the other hand, no oscillations are visible for the case of the DF $f_{eq,||}$ (red-solid curves), 
in which the small departure from the initial configuration (about $10^{-4}\%$), is presumably due to the numerical error in the calculation 
of the velocity moments of $f_{eq,||}$. 

\subsection{Perpendicular case}

For this case, we discretized the four dimensional phase space through $N_x=512$ grid points in the one-dimensional spatial domain and 
$N_v=141$ grid points in each velocity direction, while $v_{max}=7 v_{th,p}$. Two simulations have been performed to compare the SM DF 
$f_{SM}$ (S3) and the equilibrium DF $f_{eq,\perp}$ (S4), while the background magnetic field is ${\bf B}_0=B_0 {\bf e}_z$ ($B_0=1$ in 
scaled units) and the ion plasma beta $\beta_p=2v_{th,p}^2/V_{_A}^2=4$. In both simulations the system dynamics is followed up to a time 
$t_{max}=40\Omega_{cp}^{-1}$ and no perturbations are introduced. The initial electric field considered for these simulations is the one 
given by:
\begin{equation}\label{Enum}
 E(x) = E_0\left[ 1 - \tanh{\left( \frac{x-L/4}{\Delta x} \right)} + \tanh{\left( \frac{x-3L/4}{\Delta x} \right)} \right]
\end{equation}
with $E_0=1$ (in scaled units), $\Delta x=2.5 d_p$ and $L=100 d_p$. Note that the shear has been mirrored to hold periodic boundary 
conditions.

\begin{figure*}[!htb]
\epsfxsize=12.5cm \centerline{\epsffile{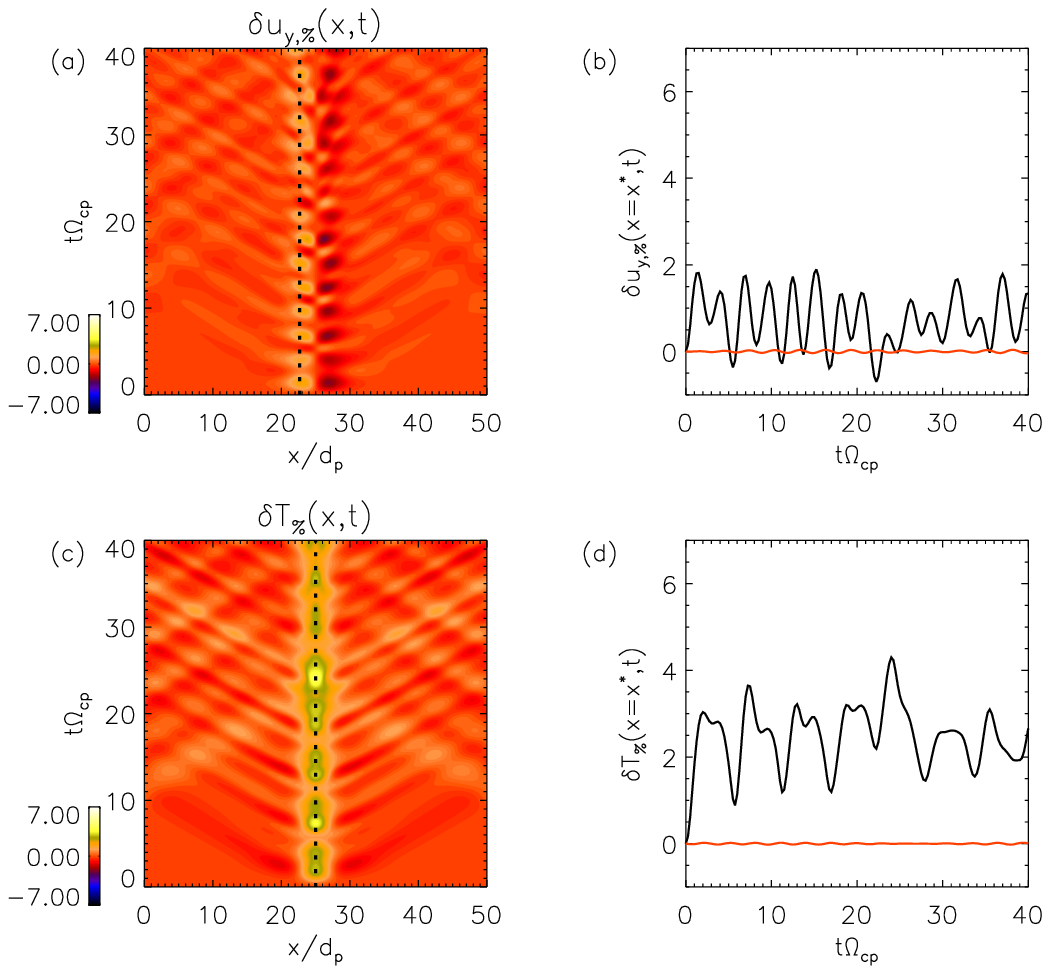}}   
\caption{Top: contour plot of $\delta u_{y,\%}(x,t)$ (a) and the temporal profile of $\delta u_{y,\%}(x=x^*,t)$ (b), being $x=x^*$ 
indicated in panel (a) with the black dashed line. Bottom: contour plot of $\delta T_{\%}(x,t)$ (c) and the temporal profile of $\delta 
T_{\%}(x=x^*,t)$ (d), being $x=x^*$ indicated in panel (c) with the black dashed line.}
\label{Fig:eqBperpU}
\end{figure*}

As expected, the simulation S3 indicates that the initial distribution SM DF is not an equilibrium. However, with respect to the 
parallel case, its velocity moments does not display an oscillatory behaviour but some propagating structure is also recovered. Left column 
of Fig. \ref{Fig:eqBperpU} displays the contour plots of $\delta u_{y,\%}(x,t) = (u_y(x,t)-u_y(x,0))/U_0\times100$ [panel (a)] and $\delta 
T_{\%}(x,t)=(T(x,t)-T(x,0))/T^{\infty}\times100$ [panel (c)] in the $(x,t)$ plane, for the simulation S3, where the $x$ range has 
been set to focus on the left half of the spatial box and where $T^{\infty} = v_{th,p}^2= \beta_p/2 = 2$ in dimensionless units. 
Disturbances from the equilibrium, (about $1-2\%$) are recovered in both $u_y$ and $T$, mainly localized around the shear but also showing 
a propagating structure.

On the other hand, in the simulations S4 with the equlibrium DF, the system remains at equilibrium (oscillations significantly smaller 
than in S3 are in fact recovered in S4). To better point out the differences between S3 and S4, we considered the temporal profiles of 
$\delta u_{y,\%}(x,t)$ and $\delta T_{\%}(x,t)$, evaluated at a fixed spatial position $x=x^*$. The point $x^*$ corresponds to the spatial 
point where each quantity 
exhibits the largest amplitude oscillations in simulation S3 (vertical black-dashed lines in panels (a) and (c) of Fig. 
\ref{Fig:eqBperpU}), 
with respect to the initial condition, i.e. $\delta u_{y,\%}(x^*,t)=\max_x\{\delta u_{y,\%}(x,t)\}$ and $\delta 
T_{\%}(x^*,t)=\max_x\{\delta T_{\%}(x,t)\}$. These temporal profiles are reported in panels (b) and (d) of Figure \ref{Fig:eqBperpU}, as 
black and red curves, respectively. Here, one can realize that much smaller oscillations in the signals are recovered in the case of the 
DF $f_{eq,\perp}$ (red), as compared to the case of the SMDF (black).

\begin{figure}[!htb]
\epsfxsize=7cm \centerline{\epsffile{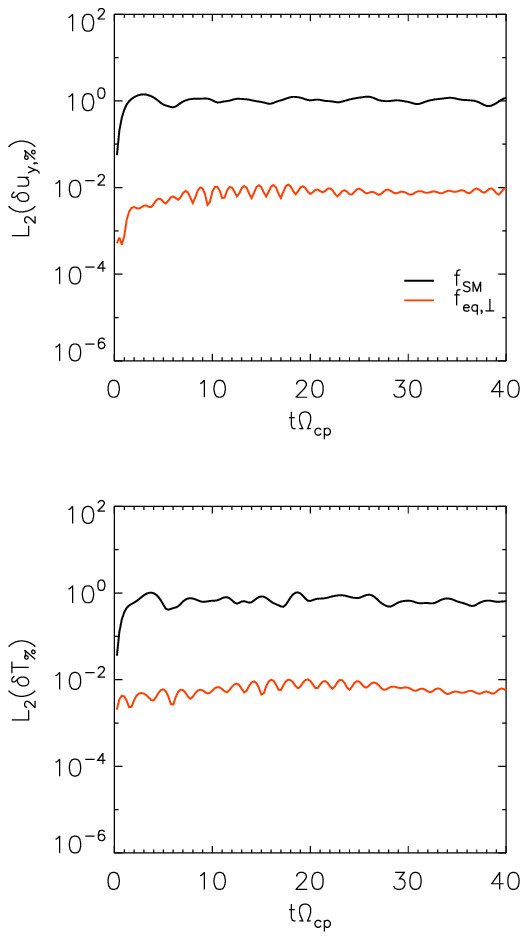}}   
\caption{Temporal evolution of $L_2(\delta u_{y,\%})$ (top row) and $L_2(\delta T_{\%})$ (bottom row). In each panel the black line 
refer to the $f_{SM}$ DF, while the red line to the $f_{eq,\perp}$ DF. }
\label{Fig:normeBperpU}
\end{figure}

To further characterize the differences between the two cases S3 and S4, we also computed the $L_2$ norm of $\delta u_{y,\%}(x,t)$ and 
$\delta T_{\%}(x,t)$, defined as $L_2(g(x,t)) = \sqrt{\int (g-g_0)^2 dx / L} $, being $g$ a generic function and $g_0=g(x,0)$. 
Clearly $L_2(g(x,t))$ is a function of time $t$. In Figure \ref{Fig:normeBperpU}, we show, in semi-logarithmic plot, the temporal evolution 
of $L_2(\delta u_{y,\%})$ (top row) and $L_2(\delta T_{\%})$ (bottom row). As it can be appreciated from the plots in this figure, 
significant departures from the initial configuration are present in the case of the SMDF (black-solid curves), confirming that 
this distribution is not a HVM equilibrium. On the other hand, in the case of the DF $f_{eq,\perp}$ (red-solid 
curves) much small departures from the initial configuration (about $10^{-2}\%$) are observed, presumably due to the numerical error in the 
calculation of the velocity moments of $f_{eq,\perp}$.

\section{Summary and conclusions}

In this paper, we have derived exact solutions for the system of Hybrid Vlasov-Maxwell equations which represents a stationary 
shearing flow with a uniform magnetic field directed either parallel or perpendicular to the plasma bulk velocity. Plasmas supporting 
shearing flows are found in many situations and a kinetic description is necessary whenever the shear width is of the order of kinetic 
scales, like, for instance, in the case of the Earth's 
magnetopause \cite{sckopke81,chen93,kivelson95,faganello17}. 

The interest of building up stationary solutions can be related to the problem of describing the propagation and evolution of waves 
in a plasma with a stable shearing flow. The interaction 
between waves and the background inhomogeneity associated with the shearing flow moves the wave energy towards small scales, where
kinetic effects are more effective. Moreover, the presence of a shearing flow can generate wave coupling, with an energy transfer among 
different wave modes. It is clear that, in order to properly study wave propagation, it is necessary that the background structure 
remains stationary; otherwise, a time evolution intrinsic of the background state would superpose to waves, making difficult to 
single out the wave contribution in the overall time evolution.
Another possible application of exact shearing flow solutions can be found in the study of the Kelvin-Helmoholtz instability, which takes 
place in unstable shearing velocity configurations. In fact, though the turbulent stage following the instability saturation should be 
quite insensitive to the details of the initial state, only a stationary unperturbed configuration allows to  
properly describe the linear stage of the instability. 
Therefore, in both cases an exact stationary distribution function is preferable to the simpler shifted Maxwellian DF.

Stationary solutions in various configurations have been described in previous studies of the fully 
kinetic case, i.e., involving the full set of ion and electron Vlasov-Maxwell equations. However, the fully kinetic treatment is 
quite complex and such solutions have rarely been employed in numerical simulations like, for instance, in investigations of the KH instability. In this respect, the set of Hybrid Vlasov-Maxwell equations represents a good compromise, because it correctly describes 
a plasma at scales of the order of or larger than the ion scales, but avoiding the complexity of a fully kinetic treatment. 
In the framework of Vlasov-Maxwell equations, Cerri et al. \cite{cerri13} have developed a method to derive approximately stationary 
solutions. The solutions presented here situate in the same framework, but have the advantage to be exactly stationary. 

The starting point of our derivation are previous studies where stationary DFs are derived in the fully kinetic framework (Refs. 
\cite{roytershteyn08,ganguli88}) which we revisited and adapted to the hybrid Vlasov-Maxwell description. In particular, we have examined 
the special cases in which the magnetic field is uniform and either parallel or perpendicular to the bulk velocity. In the former 
case the stationary solution have a simple analytical form which can be directly used in numerical simulations; in the latter case, 
the explicit construction of the distribution function is obtained through a simple numerical procedure which integrates particle orbits 
throughout the relevant phase space. In the case of parallel ${\bf B}$ an isothermal electron fluid have been assumed. In contrast, 
in the case of perpendicular ${\bf B}$ a nonuniform electron pressure $p_e$ is necessary in order to satisfy the generalized Ohm's law. 
As a consequence, an adiabatic equation for the electron fluid has been added to the set of equations. This aspect represents a novelty 
for the hybrid Vlasov-Maxwell approach, in which an isothermal electron fluid has been routinely assumed. 

The main properties of these solutions have been examined, calculating the associated profiles of bulk velocity, temperatures and heat flux. 
In the shear region, the ion distribution functions are distorted with respect to shifted Maxwellians, with stronger distortions 
for more localized shears. In particular, marked anisotropy and agyrotropy in the ion temperature are generated, and none of the DF 
principal axes is aligned to ${\bf B}$. Moreover, a non-vanishing heat flux is present, directed in the plane perpendicular 
to the inhomogeneity direction $x$. We found that the width of the velocity shear cannot be smaller than ion Larmor radius; this can be 
justified by considering the ion gyromotion which mixes up single particle velocities on the scale of the Larmor radius.
 
The HVM code \cite{valentini07} has been employed to verify to what extent the derived configurations remain stationary when used 
as initial conditions in numerical simulations. The time behaviour has been compared with that obtained in the case of a shifted Maxwellian 
distribution function. We found that in the case of our solutions the deviation from the initial condition remains much smaller 
(two orders of magnitude for perpedicular ${\bf B}$ and more than three orders of magnitude for parallel ${\bf B}$) than in the case 
of the shifted Maxwellian. The small deviation from exact stationarity of the former case are probably due to numerical errors in 
the HVM code and, for perpendicular ${\bf B}$, also in the procedure integrating particle orbits.

We are planning to use these results for studying the problem of Alfv\'en wave evolution in a shearing flow plasma. Moreover, we are 
currently working to extend them to the case of an obliquely-directed magnetic field, a configuration commonly observed in the 
Earth's magnetosphere.

\section*{Acknowledgement}
This work has been supported by the Agenzia Spaziale Italiana under the contract ASI-INAF 2015-039-R.O ``Missione M4 di ESA: Partecipazione 
Italiana alla fase di assessment della missione THOR''.

\appendix
\section{Anisotropy evaluation at the center of the shear in the parallel case}
\label{app:anisev}

In this Appendix, we analytically calculate, in the case of the shear parallel to the background magnetic field, the variance matrix for 
the equilibrium DF $f_{eq,||}$ of Eq. \ref{DF2}. We focus on the case where $U$ represents a shear layer across which the bulk velocity 
varies from a value $-U_0$ to $U_0$. In the spatial positions far from the velocity shear $f_{eq,||}$ reduces to a shifted Maxwellian, then 
the variance matrix becomes diagonal $\sigma_{ij}^{\infty} = v_{th,p}^2 \delta_{ij}$ and 
$T_{0}^{\infty}=T_{||}^{\infty}=T_{\perp}^{\infty}$, where the upper index ``${\infty}$" identifies values calculated far from the shear 
layer. We focus the center $x=0$ of a symmetric shear layer, i.e. $U(k_2)$ is an odd function of $k_2$, where we expect to find the 
strongest departures from a Maxwellian. In this case, $u_y(x=0)=0$, while the variance matrix is:
\begin{equation}\label{varmat2}
\sigma_{ij}(x=0)= \frac{1}{(2\pi )^{3/2} v_{th,p}^3} \int v_i v_j \exp \left\{ -\frac{1}{2 v_{th,p}^2} \left[ v_x^2 + \left( v_y + U\left( 
\frac{v_z}{\Omega_{cp}} \right) \right)^2 +v_z^2 \right] \right\} d^3{\bf v}
\end{equation}
In the large scale limit $\Delta x \gg \rho_{_{Lp}}$ ($\Delta v \gg v_{th,p}$) and by considering that the typical value for the velocity 
$v_z$ is $v_{th,p}$, we can retain the first-order Taylor expansion term of $U(v_z/\Omega_{cp})$: $ U(v_z/\Omega_{cp}) \simeq 
\omega_0/\Omega_{cp} v_z $, being $\omega_0 \equiv \left. \frac{dU}{dk_2} \right|_{k_2=0}$

Within this approximation, all the integrals involved in Eq. (\ref{varmat2}) can be easily calculated and the resulting form is:
\begin{equation}\label{varmat3}
\sigma(x=0) = 
\left[ \begin{array}{ccc}
v_{th,p}^2 & 0 & 0 \\
0 & \displaystyle{\left( 1+\frac{\omega_0^2}{\Omega_{cp}^2}\right) v_{th,p}^2} & 
\displaystyle{-\frac{\omega_0}{\Omega_{cp}} v_{th,p}^2} \\
0 & \displaystyle{-\frac{\omega_0}{\Omega_{cp}} v_{th,p}^2} & v_{th,p}^2 
\end{array} \right]
\end{equation}

Diagonalizing $\sigma(x=0)$ implies to rotate the DF into the minimum variance frame (MVF), The eigenvalues of $\sigma(x=0)$, corresponding 
to the temperatures in the MVF, are $\lambda^{(3)} < \lambda^{(2)} < \lambda^{(1)}$, whose explicit expressions are:
\begin{eqnarray}
\lambda^{(3)} &=& v_{th,p}^2 \left( 1- \frac{\omega_0}{\Omega_{cp}} \sqrt{1 + \frac{\omega_0^2}{4\Omega_{cp}^2}} + 
\frac{\omega_0^2}{2\Omega_{cp}^2} \right) \nonumber \\
\lambda^{(2)} &=& v_{th,p}^2 \label{lambda} \\
\lambda^{(1)} &=& v_{th,p}^2 \left( 1+ \frac{\omega_0}{\Omega_{cp}} \sqrt{1 + \frac{\omega_0^2}{4\Omega_{cp}^2}} + 
\frac{\omega_0^2}{2\Omega_{cp}^2} \right) \nonumber
\end{eqnarray}
The corresponding eigenvectors are given by:
\begin{eqnarray}
\xi^{(3)} &=& \left( \sqrt{1 + \frac{\omega_0^2}{4\Omega_{cp}^2}}-\frac{\omega_0}{2\Omega_{cp}} \right) {\bf e}_y + {\bf e}_z \nonumber \\
\xi^{(2)} &=& {\bf e}_x \label{xi}\\
\xi^{(1)} &=& -\left( \sqrt{1 + \frac{\omega_0^2}{4\Omega_{cp}^2}}+\frac{\omega_0}{2\Omega_{cp}} \right) {\bf e}_y + {\bf e}_z \nonumber
\end{eqnarray}
From the above expressions we can deduce the following informations: 

\noindent (i) At $x=0$, $\sigma_{xx} = \sigma_{zz}$, therefore the two temperatures in the directions orthogonal to ${\bf B}$ are equal, 
i.e. the DF is gyrotropic in the LBF at $x=0$.

\noindent (ii)
By comparing $\sigma_{ij}^{\infty}$ and $\sigma_{ij}(x=0)$ in Eq. (\ref{varmat3}), we note that, at $x=0$, $\sigma_{xx}$ and $\sigma_{zz}$ 
keep the same value of $\sigma_{ij}^{\infty}$. Therefore, at the center of the shear, the perpendicular proton temperature remains constant 
($T_{\perp}(x)=T_{\perp}^{\infty}$), while the parallel temperature increases ($T_{||}(x=0) \ge T_{||}^{\infty}$). As a consequence, 
at $x=0$ the perpendicular to parallel proton temperature ratio is 
\begin{equation}\label{Tratio}
\frac{T_{\perp}(x=0)}{T_{||}(x=0)} = \frac{1}{1+\omega_0^2/(2\Omega_{cp}^2)} < 1
\end{equation}
i.e., the proton parallel temperature is larger than the perpendicular one.

\noindent (iii)
The eigenvectors $\xi^{(m)}$ ($m=1,2,3$) give the directions of the principal axes of the DF in the velocity space. Far from the shear 
layer the DF is isotropic and the directions of principal axes are arbitrary. At $x=0$, the principal axis $\xi^{(2)}$ is in the $v_x$ 
direction, which corresponds to the direction of spatial variation of the bulk velocity ${\bf u}$, while the other two principal axes 
$\xi^{(1)}$ and $\xi^{(3)}$ are in the $v_yv_z$ plane. The angle $\gamma$ between $\xi^{(1)}$ and the $v_y$ axis, which gives the 
directions 
corresponding to the maximum width of the DF, is:
\begin{equation}\label{tantheta}
\tan \gamma = -\frac{1}{\displaystyle{\sqrt{1+\frac{\omega_0^2}{4\Omega_{cp}^2}}+\frac{\omega_0}{2\Omega_{cp}}}}
\end{equation}
In the limit of slowly varying bulk velocity ($\omega_0 \ll \Omega_{cp}$), we get $\gamma \simeq -45^\circ$. Finally, it is worth noting 
that, since the three eigenvalues are all different at $x=0$, the DF is not gyrotropic with respect to any of the three principal axes in 
the MVF. 

It is interesting to extend the results illustrated above through the numerical evaluation of the temperature in the MVF and in the LBF, 
for several values of the shear width $\Delta x$, where the shear function $U(x)$ is given by Eq. (\ref{shear0}). Table \ref{Tab:param} 
reports the values of $\lambda^{(i)}$ (with $i=1,2,3$), $T_{||}$, $T_{\perp}$ and $\gamma$, at the center of the shear $x=0$. As 
expected, in the large scale limit $\Delta x \gg \rho_{_{Lp}}$, the results obtained numerically are in good accordance with the 
analytical predictions. By decreasing the width of the shear function $\Delta x$, the analytical calculations tend to diverge from  
numerical evaluations. Note also that, as the shear width $\Delta x$ becomes smaller, stronger anisotropies in the LBF (i.e. bigger 
$T_{||}$) and in the MVF (i.e. larger ratios between the eigenvalues $\lambda^{(i)}$) are recovered at the center of shear.

\section{Stationary solution in the local approximation in the perpendicular case}
\label{app:LAeq}

In the present Appendix, we report the evaluation of the equilibrium DF, in the perpendicular case, within the so called ''local 
approximation``, i.e. in the simplified case of linearly growing electric field. The single particle motion is calculated by solving Eqs. 
(\ref{motioneq}), in the particular case in which the electric field is linear $E(x)=E_0+\alpha_0(x-x_0)$, with $E_0$ and $\alpha_0$ 
constant. This profile for the electric field is not fully realistic, since $|E(x)|$ grows without limit for increasing $|x-x_0|$. 
However, it can be considered as a local approximation of an electric profile $E(x)$ around a given position $x_0$, being 
$\alpha_0=(dE/dx)(x_0)$. 

In this case, Eq. (\ref{eqx}) reads:
\begin{equation}\label{harmosc}
\frac{d^2 x}{dt^2} + \omega^2 x = \frac{e}{m_p}\left( E_0 - \alpha_0 x_0\right) + \Omega_{cp} W_0
\end{equation}
where $\omega^2 = \Omega_{cp}^2 - e \alpha_0/m_p$. If $\Omega_{cp}^2 > e\alpha_0/m_p$, Eq. (\ref{harmosc}) describes an harmonic 
oscillator of solution:
\begin{equation}\label{xlin0}
x(t) = R_0 \sin \left( \omega t + \varphi \right) + \frac{1}{\omega^2} \left[ \frac{e}{m_p} \left( E_0-\alpha_0 x_0 \right) +\Omega_{cp} 
W_0 \right]
\end{equation}
with $R_0$ and $\varphi$ the amplitude and the phase of the motion, respectively. The constant term in Eq. (\ref{xlin0}) 
represents the time-averaged $x$ position, i.e. the guiding center position $x_c$:
\begin{equation}\label{xclin}
x_c  = \frac{1}{\omega^2} \left[ \frac{e}{m_p} \left( E_0-\alpha_0 x_0 \right) +\Omega_{cp} W_0 \right]
\end{equation}
and then
\begin{equation}
 \begin{cases}
 &  x(t) = R_0 \sin \left( \omega t + \varphi \right) + x_c \\
 & v_x(t)= R_0 \omega \cos \left( \omega t + \varphi \right) \\
 & v_y(t)= -\Omega_{cp} x(t)+W_0 = -R_0 \Omega_{cp} \sin \left( \omega t + \varphi \right) + v_{cy} 
\end{cases}
\label{xvxvylin}
\end{equation}
where $v_{cy}  = \langle v_y(t) \rangle_t = -\Omega_{cp} x_c + W_0 $. Using Eq. (\ref{xclin}) into the $v_{cy}$ expression, one obtains:
\begin{equation}\label{vyclin2}
v_{cy} = -\frac{e}{m_p \Omega_{cp}^2} \left( \Omega_{cp} E_0 + \alpha_0 v_{0y} \right)
\end{equation}
where $v_{0y}=-\Omega_{cp} x_0+W_0$ is the particle streamwise velocity component at the position $x=x_0$. Note also that 
$v_{0y}=v_{cy}+\Omega_{cp}(x_c-x_0)$. By using these last expressions, one obtains:
\begin{equation}\label{vyclin3}
v_{cy}=-\frac{e}{m_p \Omega_{cp}} \left[ E_0+\alpha_0 \left( x_c - x_0 \right)\right] = -c \frac{E(x_c)}{B}
\end{equation}
Hence, the guiding center moves along $y$ with the local ${\bf E} \times {\bf B}$ drift velocity calculated at the guiding center position 
$x_c$. Then, in the local approximation case the particle orbit in the $v_x v_y$ plane is an ellipse elongated along $v_x$ ($v_y$) for 
$\alpha_0 <0$ ($0 < \alpha_0 < m_p \Omega_{cp}^2/e$), while it reduces to a circle in the case of uniform electric field $\alpha_0=0$. In 
the case $\alpha_0 \ge m_p \Omega_{cp}^2/e$, $x(t)$ increases linearly or exponentially in time, causing the breakdown of the local 
approximation, as already noticed in Ref. \cite{nishikawa88}. 


Note that, if $\alpha_0=0$ (i.e. $\omega^2=\Omega_{cp}^2$), then $v_x^2+(v_y-v_{cy})^2=R_0^2 \omega^2$. Indicating by 
$t_n=(n\pi-\phi)/\omega$ ($n$ integer) an instant of time when $x(t_n)=x_c$, then $v_x(x=x_c)=v_x(t=t_n)=\pm R_0\omega$ [Eqs. 
(\ref{xvxvylin})]. Hence, $v_x^2+(v_y-v_{cy})^2=\left[ v_x(x=x_c)\right]^2$. The same argument show that $v_x(x=x_c)=\pm R_0\omega$ also 
when the electric field is non-uniform ($\alpha_0 \ne 0$). Then, the energy $\mathcal{E}_0$ [Eq. (\ref{toten2})] is:
\begin{equation}\label{elin}
\mathcal{E}_0=\frac{m_p}{2}  \left( R_0^2 \omega^2 + v_z^2 \right)
\end{equation}
In order to explicitly write the form for $f_{eq,\perp}(x,{\bf v})$ [Eq (\ref{fgen})], we manipulate Eq. (\ref{elin}) by using Eqs. 
(\ref{xvxvylin}) and Eq. (\ref{vyclin3}) and by expressing $x_c$ in terms of the particle position and velocity through Eqs. 
(\ref{xvxvylin}):
\begin{equation}\label{proc3}
\mathcal{E}_0 = \frac{m_p}{2} \left[ v_x^2+\frac{\omega^2}{\left( \Omega_{cp}-c\alpha_0/B \right)^2} \left( v_y + \frac{c}{B} E(x) 
\right)^2 + v_z^2 \right]
\end{equation}
This expression, which is the argument in the exponential of the DF $f_{eq,\perp}$, suggests that $f_{eq,\perp}$ is a shifted bi-Maxwellian 
with different temperatures $T_u=T_y$ and $T_{\perp u}=T_{xz}$ parallel to and in the plane perpendicular to the bulk flow, respectively. 
The temperature ratio is $T_u/T_{\perp u}=(\Omega_{cp}-c\alpha_0/B)^2/\omega^2$; using the explicit expression for $\omega$ we find:
\begin{equation}\label{TratioBperpU}
\frac{T_u}{T_{\perp u}}=1-\frac{1}{\Omega_{cp}} \frac{c \alpha_0}{B}=1-\frac{e \alpha_0}{m_p \Omega_{cp}^2}
\end{equation}
Therefore, $T_u > T_{\perp u}$ ($T_u < T_{\perp u}$) when $E(x)$ decreases (increases) with increasing $x$. Note that, in order to 
have a positive temperature $T_u$, the spatial derivative of the electric field has an upper limit: $\alpha_0 < m_p \Omega_{cp}^2/e$. 
This condition is the same which avoids the breakdown of the local approximation in the single ion dynamics, as found in the previous 
section. Of course, for a uniform electric field ($\alpha_0=0$), we have $T_u = T_{\perp u}$.

By requiring that the uniform density $n^{LA} (x)=n_0$, it can be easily shown that:
\begin{equation}\label{DFLAnorm}
f_{eq,\perp}^{LA}(x,{\bf v}) = \frac{n_0}{(2\pi)^{3/2} v_{th,p}^3} \left( \frac{T_{\perp u}}{T_u} \right)^{1/2}
e^{\displaystyle{-\frac{1}{2v_{th,p}^2} \left\{ v_x^2+
\frac{T_{\perp u}}{T_u} \left[ v_y + \frac{c}{B} E(x) \right]^2 + v_z^2 \right\} }}
\end{equation}
which depends on two arbitrary constants $n_0$ and $v_{th,p}$, while the ratio $T_u/T_{\perp u}$ is given by the equation 
(\ref{TratioBperpU}).

The bulk velocity associated with the DF is ${\bf u}^{LA}=\int {\bf v}\, f_{eq,\perp}^{LA}\, d^3{\bf v}/n^{LA}$. The bulk velocity 
components $u^{LA}_x$ and $u^{LA}_z$ are both vanishing, while the component $u^{LA}_y= - c E(x)/B$, indicating that the bulk velocity 
coincides with the local ${\bf E} \times {\bf B}$ drift velocity.

The considered DF is an exact stationary solution of the HVM equations Eqs. (\ref{vlasoveq}--\ref{E}), with Eq. (\ref{eadiab}) for the 
pressure closure. The DF $f_{eq,\perp}^{LA}$ is a stationary solution of the Vlasov equation [Eq. (\ref{vlasoveq})], because it is a 
function only of the motion constant. The electric field profile is linear and its profile is correctly given by the ${\bf u} \times {\bf 
B}$ term in Eq. (\ref{E}) (then the stationarity holds for an electron pressure $p_e$ constant and uniform). Since the electric field is 
irrotational, the magnetic field is stationary. Therefore, the considered configuration is an exact stationary solution of the whole set of 
the HVM equations.



\newpage

\clearpage
\begin{table}[!b]
\begin{center}
\begin{tabular}{|cc|cccccc|}
\tableline\tableline
$\Delta_x/d_p$ & Evaluation & $\lambda_1$ & $\lambda_2$ &  $\lambda_3$ & $T_{\parallel}$ & $T_{\perp}$ & $\gamma$ \\
\tableline
$25$ & Analytical & $2.16$ & $2.00$ & $1.85$ & 2.01 & 2.00 & -43.9 \\
$25$ & Numerical & $2.16$ & $2.00$ & 1.85 & 2.01 & 2.00 & -43.9 \\
\tableline
$2.5$ & Analytical  & $4.36$ & $2.00$ & $9.17\times10^{-1}$  & 3.28 & 2.00 & -34.1 \\
$2.5$ & Numerical & $3.75$ & $2.00$ & 1.08 & 2.82 & 2.00 & -36.0 \\
\tableline
$0.25$ & Analytical  & $1.32\times10^2$ & 2.00 & $3.03\times10^{-2}$ & $1.30\times10^{2}$ & 2.00 & -7.02 \\
$0.25$ & Numerical & $6.53$ & 2.00 & $9.09\times10^{-1}$ & 5.44 & 2.00 & -26.1 \\
\tableline
\end{tabular}
\end{center}
\caption{Temperatures and characteristic angle of the equilibrium distribution function $f_{eq,||}$.}.
\label{Tab:param}
\end{table}

\end{document}